\documentclass[11pt]{article}
\pdfoutput=1

\usepackage{jheppub}
\usepackage{amssymb,amsfonts,amsmath,amsthm,lineno}
\usepackage{comment}
\usepackage{enumerate}
\usepackage{mathrsfs}
\usepackage{bbold}
\usepackage{tikz}
\usetikzlibrary{shapes.geometric, arrows}
\usepackage{hyperref}
\usepackage{slashed}
\usepackage{color}
\usepackage{empheq}
\usepackage{soul}
\usepackage{cancel}
\usepackage{placeins}
\usepackage{xargs}
\usepackage{tikz-cd}

\makeatletter
\newcommand{\Vast}{\bBigg@{4.75}}
\makeatother

\newcommand{\be}{\begin{equation}}
\newcommand{\ee}{\end{equation}}
\newcommand{\bea}{\begin{eqnarray}}
\newcommand{\eea}{\end{eqnarray}}

\newcommand{\CC}{\mathcal{C}}
\newcommand{\CD}{\mathcal{D}}
\newcommand{\CE}{\mathcal{E}}

\newcommand{\CL}{\mathcal{L}}
\newcommand{\CN}{\mathcal{N}}

\newcommand{\CO}{\mathcal{O}}

\newcommand{\CQ}{\mathcal{Q}}

\newcommand{\CZ}{\mathcal{Z}}

\newcommand{\lr}{\left (}
\newcommand{\rr}{\right )}

\newcommand{\p}{\partial}

\renewcommand{\tilde}[1]{\widetilde{#1}}
\newcommand{\tr}{\text{tr}}

\makeatletter
\renewcommand{\@seccntformat}[1]{\csname the#1\endcsname.\,\,}
\makeatother
\allowdisplaybreaks

\usetikzlibrary{shapes,arrows,chains}
\usetikzlibrary{decorations.markings}
\usetikzlibrary{decorations.pathmorphing}
\usetikzlibrary{shapes.multipart}
\tikzset{snake it/.style={decorate, decoration=snake}}

\usepackage{tikz-cd}
\usetikzlibrary{cd,decorations.markings, arrows.meta}

\newcommand\qt\tau

\newcommand{\dd}{\mathrm{d}}

\let \savenumberline \numberline
\def \numberline#1{\savenumberline{#1.}}

\makeatletter
\def\@fpheader{\relax}
\makeatother

\def\bea{\begin{eqnarray}}
\def\eea{\end{eqnarray}}

\newcommand\blfootnote[1]{%
\begingroup 
\renewcommand\thefootnote{}\footnote{#1}%
\addtocounter{footnote}{-1}%
\endgroup 
}

\title{
Heterotic String Sigma Models: Discrete Light Cone Quantization and Its  Current-Current Deformation
}
\author[a]{Eric A. Bergshoeff,}
\author[b]{Kevin T. Grosvenor,}
\author[c]{Luca Romano,}
\author[d]{Ziqi Yan
\blfootnote{*The authors are ordered purely alphabetically and should all be viewed as the co-first authors.}}
\emailAdd{e.a.bergshoeff@rug.nl}
\emailAdd{kevinqg1@gmail.com}
\emailAdd{lucaromano2607@gmail.com}
\emailAdd{ziqi.yan@su.se}
\affiliation[a]{Van Swinderen Institute, University of Groningen \\ Nijenborgh 3, 9747 AG Groningen, The Netherlands \smallskip
}
\affiliation[b]{National Institute of Physics, University of the Philippines \\
Diliman, Quezon City 1101, Philippines \smallskip}
\affiliation[c]{
Departamento de Electromagnetismo y Electronica, Universidad de Murcia, \\
Campus de Espinardo, 30100 Murcia, Spain
}
\affiliation[d]{
Nordita, KTH Royal Institute of Technology and Stockholm University\\
Hannes Alfv\'{e}ns v\"{a}g 12, SE-106 91 Stockholm, Sweden 
}
\preprint{NORDITA 2025-024}
\abstract{We propose a two-dimensional superstring sigma model that defines a self-contained corner of heterotic string theory, whose second quantization is heterotic matrix string theory. This worldsheet theory arises from a BPS decoupling limit that zooms in on the heterotic string, under which the target space geometry becomes non-Lorentzian. This construction generalizes the Gomis-Ooguri formulation of non-relativistic string theory to the heterotic case. We show how such a worldsheet theory provides a first-principles definition of the heterotic string in the discrete light cone quantization via T-duality. By turning on a current-current deformation akin to a $T\bar{T}$ deformation, the conventional heterotic string theory with a Lorentzian target space is recovered. We analyze the gauge and gravitational anomalies with respect to the lowest-order quantum corrections in the sigma model and show how the worldsheet theory is consistent with non-Lorentzian supergravity in the target space.}

\begin{document}

\maketitle

\section{Introduction}

The discovery of heterotic string theory was one of the important achievements that ignited the first superstring revolution~\cite{Green:1984sg, Gross:1984dd}. In particular, the $E_8 \times E_8$ heterotic string appeared to be more promising than other superstring theories for phenomenological applications, as the Standard Model gauge group nicely fits inside the $E_8$ gauge group. It was then realized that all the distinct superstring theories are dual corners of a hypothetical M-theory in 11D~\cite{Hull:1994ys, Witten:1995ex}, whose dynamics was famously conjectured by Banks, Fischler, Shenker, and Susskind (BFSS) to arise non-perturbatively from a large $N$ limit of the matrix quantum mechanics of $N$ non-relativistic D-particles in type IIA superstring theory~\cite{deWit:1988wri, Banks:1996vh}. 

An alternative perspective of the large $N$ limit is via the holographic principle, where a (conformally) AdS bulk geometry emerges. In particular, BFSS matrix quantum mechanics is related to $\CN = 4$ SYM via a T-duality transformation, and the latter corresponds to gravity on AdS${}_5 \times S^5$~\cite{Maldacena:1997re}. The near-horizon limit leading to this Lorentzian bulk geometry corresponds at the asymptotic infinity to a BPS decoupling limit of type IIB superstring theory.
It was recently observed that this BPS decoupling limit in the asymptotic infinity regime leads to a 10D spacetime that is non-Lorentzian, which means that a foliation structure is developed and there is \emph{no} 10D metric description. This observation led to new insights showing that non-Lorentzian geometry provides an important tool for mapping out matrix theories and holographic dualities in string theory~\cite{Blair:2024aqz}.\,\footnote{See~\cite{Lambert:2024uue, Lambert:2024yjk} for related developments on non-Lorentzian holography and~\cite{Danielsson:2000mu, Avila:2023aey, Guijosa:2023qym} for previous insights from the dual non-relativistic string side. More recent works along these lines have appeared in~\cite{Lambert:2024ncn, Guijosa:2025mwh, Harmark:2025ikv, Blair:2025prd, Blair:2025ewa}. See also \cite{Fontanella:2024rvn, Fontanella:2024kyl, Fontanella:2024hgv} for another proposal of non-Lorentzian holography where the coordinates are treated differently.}

In this paper, we will extend this non-Lorentzian perspective developed for type II superstring theory to the heterotic case. In particular, we will formulate a new two-dimensional supersymmetric sigma model that acts as a key ingredient for building up this overarching framework  for the heterotic string. Our immediate motivations are twofold, including  aspects of heterotic matrix theory and supergravity, which we outline below.

\paragraph{\emph{Heterotic matrix theory.}} The surprising appearance of non-Lorentzian geometry in the AdS/CFT correspondence was realized from classifying decoupling limits of type II superstring theory that are central to matrix theory and holography~\cite{bpslimits, Blair:2023noj, Gomis:2023eav}. 
Such a decoupling limit zooms in on a background BPS configuration formed by branes (and strings), and typically leads to a 10D non-Lorentzian target space. The self-contained corners of string and M-theory from taking the decoupling limits constitute an intricate duality web that centers around M-theory in the discrete light-cone quantization (DLCQ)~\cite{Susskind:1997cw, Seiberg:1997ad, Sen:1997we}, \emph{i.e.}~M-theory on a lightlike compactification. The light excitations in DLCQ M-theory are the Kaluza-Klein states in the lightlike circle, whose dynamics is described by BFSS matrix quantum mechanics at finite $N$. Among all the BPS decoupling limits, the one zooming in on the fundamental string plays a pivotal role, which leads to the so-called non-relativistic string theory~\cite{Klebanov:2000pp, Gomis:2000bd, Danielsson:2000gi} (see~\cite{Oling:2022fft, Demulder:2023bux} for reviews of recent studies). This is a unitary and ultra-violet complete string theory with a Galilean-invariant closed string spectrum. The associated target space geometry is now non-Lorentzian, which is equipped with a codimension-two foliation structure satisfying a (string-)Galilean boost symmetry. The second quantization of the non-relativistic strings is matrix string theory described by 2D $\CN = 8$ super Yang-Mills (SYM)~\cite{Motl:1997th, Dijkgraaf:1997vv}, which is U-dual to BFSS matrix quantum mechanics on a spatial circle.\,\footnote{See related discussions in~\cite{Danielsson:2000gi, Blair:2024aqz, Harmark:2025ikv}.} 

We put forward in this paper a new sigma model describing the heterotic non-relativistic string, whose second quantization corresponds to heterotic matrix string theory
(see \emph{e.g.}~\cite{Danielsson:1996es, Kachru:1996nd, Motl:1996xx, Kim:1997uv, Lowe:1997fc, Banks:1997it, Banks:1997zs, Lowe:1997sx, Rey:1997hj, Horava:1997ns, Motl:1997tb, Krogh:1998vb, Krogh:1998rw}). 
We will show how this sigma model provides a first-principles definition of heterotic string theory in the DLCQ, which acquires interesting twists. Starting with heterotic non-relativistic string theory, a duality web analogous to the type II version in~\cite{Blair:2023noj, Gomis:2023eav} can be constructed, which may provide new insights for heterotic M(atrix)-theories and holography. 

\paragraph{\emph{Heterotic supergravity.}} On the other hand, our work in part originated from an attempt to overcome the intricate computations in heterotic supergravity by developing a string worldsheet method. 
We will apply our construction of heterotic string sigma models to provide a simple worldsheet perspective of two supergravity results. Before describing the new worldsheet insights in this paper, we first review the previous supergravity results below:
\begin{enumerate}[(1)]

\item

\emph{Occurrence of a generalized metric in heterotic T-duality via supergravity}~\cite{Bergshoeff:1995as}.
Supergravity has been powerful in unraveling certain perturbative as well as non-perturbative properties of string theory. Notable examples are the Green-Schwarz anomaly cancellation \cite{Green:1984sg} using the Chern-Simons term of 10D supergravity coupled to Yang-Mills (YM)~\cite{Chapline:1982ww}, D-branes as soliton solutions of IIA/IIB supergravity \cite{Polchinski:1995mt} and 11D supergravity as a rationale for the supermembrane~\cite{Bergshoeff:1987cm}. 
Another application of supergravity is the  derivation of the explicit T-duality rules that map solutions of IIA supergravity to solutions of IIB supergravity and \emph{vice versa}~\cite{Bergshoeff:1995as}. This T-duality requires an isometry direction in each of the two theories. The explicit T-duality rules, including the dilaton, can be derived by reducing over the isometry directions and comparing the two theories from a 9D point of view. The same technique has been applied many years ago to derive the T-duality rules of heterotic string theory, with the intriguing result that the heterotic T-duality transformations could be naturally formulated in terms of a generalized  metric~\cite{Bergshoeff:1995cg},
\begin{equation}\label{effectivem}
    G^{}_\text{MN} = g^{}_\text{MN} + B^{}_\text{MN} + \tfrac{1}{2} \, \tr \bigl( V^{}_\text{M} \, V^{}_\text{N} \bigr)\,,
\end{equation}
where $g^{}_\text{MN}\,, B^{}_\text{MN}$ and $V^{}_\text{M}$ are respectively the metric, Kalb-Ramond and YM vector field of  10D heterotic supergravity. At the time the emergence of this generalized metric in the heterotic T-duality rules was surprising and its origin was not well understood. 

\item

\emph{Construction of a non-Lorentzian limit of heterotic supergravity}~\cite{Bergshoeff:2023fcf}.
The second more recent heterotic supergravity result is related to a self-consistent non-Lorentzian limit, which is BPS in nature and zooms in on a background heterotic non-relativistic string. The explicit realization of such a limit turns out to be highly non-trivial in heterotic supergravity with YM fields~\cite{Bergshoeff:2023fcf}. 
One intricacy of this limit is inherited from the fact that the classical heterotic sigma models are anomalous. Worldsheet anomaly cancellation requires that the Kalb-Ramond two-form transform non-trivially under the YM gauge symmetry. This extra transformation of the Kalb-Ramond two-form  is key to the Green-Schwarz anomaly cancellation in the target space~\cite{Green:1984sg}.  
Consequently, the reparametrization of the background metric, Kalb-Ramond, and dilaton fields in terms of the would-be background fields in heterotic non-relativistic string theory, which defines the associated BPS decoupling limit, must involve the YM gauge fields. Without having a further clue of the precise reparametrization, this mixture with the YM gauge fields complicates the analysis considerably. After a long trial and error calculation, the prescriptions defining a consistent limit were finally found recently in~\cite{Bergshoeff:2023fcf}.

\end{enumerate}

We will show how the complicated heterotic supergravity computations can be replaced with a much simpler procedure by taking a string worldsheet perspective. 
More specifically, both the generalized metric~\eqref{effectivem} and the BPS decoupling limit zooming on the heterotic string can be understood in an elegant and simple way using the heterotic sigma models. As we will see, an explanation of the generalized metric~\eqref{effectivem} requires a careful consideration of the YM anomalies and a determination  of the 
quantum effective string action for the conventional heterotic string theory. We will then construct a quantum effective worldsheet action for the heterotic non-relativistic string from T-dualizing the DLCQ of heterotic string theory. Finally, concerning the 2nd supergravity result, we will show that the heterotic non-relativistic string limit can be derived in a systematic way by introducing a simple current-current deformation in the sigma model describing the heterotic non-relativistic string. This deformation is closely related to the idea of $T\bar{T}$ deformation that drives a flow from non-relativistic to relativistic string theory. This result generalizes the previous studies in~\cite{Blair:2020ops, Yan:2021lbe, Blair:2024aqz}.

\vspace{3mm}
 
This paper is organized as follows. In Section~\ref{sec:hssmdlcq}, we develop the central ingredients of the heterotic non-relativistic string. In Section~\ref{sec:aolea} we discuss the path integral for the conventional heterotic string that is explicitly gauge invariant with respect to the lowest-order in $\alpha'$. In Section~\ref{sec:tdt} we use the associated quantum effective sigma model to give a simple derivation of the heterotic T-duality transformations in terms of the generalized metric~\eqref{effectivem}. We will then consider in Section~\ref{sec:nonhstdlcq} a singular case of these T-duality transformations that underlies the DLCQ of heterotic string theory, which motivates our formulation of the heterotic non-relativistic string sigma model. This new sigma model in turn provides a first-principles definition of the DLCQ of heterotic string theory. In Section~\ref{sec:ccdbps}, we derive the associated BPS decoupling limit via a current-current deformation of the non-relativistic heterotic worldsheet theory that is closely related to a $T\bar{T}$-deformation. We will then generalize in Section~\ref{sec:satc} this limiting procedure to obtain the complete $\CN=1$ supersymmetric sigma models describing heterotic non-relativistic superstring theory and study the associated gauge and gravitational anomalies. In Section~\ref{sec:hmst}, we comment on the relation to heterotic matrix string theory and holography. Finally, we conclude our paper in Section~\ref{sec:concl}.  In Appendix~\ref{app:ss} we discuss a Stueckelberg symmetry of the heterotic non-relativistic string sigma model. In Appendix~\ref{app:tdhnrst} we have collected some results on 
 other types of T-duality transformations in heterotic non-relativistic string
sigma models than the one used in the main text.

\section{Heterotic String Sigma Model and Its DLCQ} \label{sec:hssmdlcq}

We now move on to review the construction of the classical heterotic string sigma model action, whose path integral is anomalous under both the YM and gravitational gauge transformations~\cite{Hull:1985jv, Sen:1985qt, Hull:1986xn}. We will follow the discussion and convention in~\cite{Polchinski:1998rr} here. After this review, in Section~\ref{sec:aolea}, we will propose a sigma model whose path integral is anomaly free with respect to the lowest-order $\alpha'$-corrections. We will see that this formulation naturally contains the generalized metric~\eqref{effectivem}. Subsequently, in Section~\ref{sec:tdt}, we will use this sigma model to reproduce the T-duality transformations that were originally obtained from supergravity in~\cite{Bergshoeff:1995cg}. 

In the second half of this section, we will study heterotic string theory in the discrete light-cone quantization (DLCQ). In Section~\ref{sec:nonhstdlcq}, we start with a brief review of the standard Gomis-Ooguri sigma model~\cite{Gomis:2000bd} and its T-dual relation to type II string theory in the DLCQ~\cite{Bergshoeff:2018yvt}. Next, we analogously focus on a `singular' case of the heterotic T-duality transformation that gives rise to the DLCQ of heterotic string theory. This will lead us to the construction of the sigma model describing the heterotic non-relativistic string.  We will develop an independent treatment of the YM gauge anomaly cancellation in this theory. The complete supersymmetric sigma model for the non-relativistic heterotic string will be given in Section~\ref{sec:satc}, where the complete theory that is free of both YM and gravitational anomaly with respect to the lowest-order $\alpha'$-corrections will be formulated.

\subsection{Anomalies and Quantum Effective Action} \label{sec:aolea}

In heterotic string theory, the $\mathcal{N}=1$ worldsheet supersymmetry is realized by combining the left movers in 26D bosonic string theory with the right movers in 10D superstring theory, and it realizes YM gauge symmetries via the left-moving currents of the string, with the charges distributed democratically along the closed strings~\cite{Gross:1984dd}. We consider the Ramond–Neveu–Schwarz (RNS) formalism for the heterotic string. 
Denote the Euclidean worldsheet in terms of the complex variable $z$ and its complex conjugate $\bar{z}$\,. Introduce 10 bosonic fields $X^\text{M} (z\,, \, \bar{z})$\,, $\text{M} = 0\,, \, \cdots, \, 9$ and 32 left-moving fermions $\lambda^A (z)$\,, $A = 1\,, \, \cdots, \, 32$\,. The worldsheet $\mathcal{N} = 1$ supersymmetry is realized by introducing ten additional chiral fermions $\psi^\text{M} (\bar{z})$\,, with the supersymmetry transformations,
\begin{equation} \label{eq:stxpsi}
    \delta X^\text{M} = - \sqrt{\frac{\alpha'}{2}} \, \bar{\epsilon} \, \psi^\text{M}\,,
        \qquad%
    \delta \psi^\text{M} = \sqrt{\frac{2}{\alpha'}} \, \bar{\epsilon} \, \p_{\bar{z}} X^\text{M}\,.
\end{equation} 
Note that $\lambda^A$ is a supersymmetry singlet. 
In conformal gauge and in flat target space, the sigma model is

\begin{equation} \label{eq:ftshssm}
    S = \frac{1}{4\pi} \! \int \! \text{d}^2 z \lr \frac{2}{\alpha'} \, \partial_z X^\text{M} \, \partial_{\bar{z}} X^{}_\text{M} + \lambda^A \, \partial_{\bar{z}} \lambda^A + \psi^\text{M} \, \partial_z \psi^{}_\text{M} \rr.
\end{equation}
From the worldsheet point of view, both $\lambda^A$ and $\psi^\text{M}$ are Majorana-Weyl spinors. 
Depending on the boundary conditions imposed on $\lambda^A$\,, this action defines the SO(32) or $\text{E}_8 \times \text{E}_8$ heterotic string, with ``$A$'' the gauge index. We set $\alpha' = 2$ below and only recover its dependence via dimensional analysis when necessary. 

In order to understand the anomalies in heterotic sigma models, one needs to examine the heterotic string in general background fields, for which it is useful to invoke the superspace formalism. Define the supercoordinate $\bar{\theta}$ and the supercovariant derivative,
$D_{\bar{\theta}} = \p_{\bar{\theta}} + \bar{\theta} \, \p_{\bar{z}}$\,.
We then introduce the superfields,
\be \label{eq:sf}
    Y^\text{M} = X^\text{M} + i \, \bar{\theta} \, \psi^\text{M}\,,
        \qquad%
    \Lambda^A = \lambda^A + \bar{\theta} \, F^A\,,
\ee
where we have defined the auxiliary bosonic field $F^A$, such that the supersymmetry transformations are now supplemented with
\be \label{eq:stxpsiex}
    \delta \lambda^A = i \, \bar{\epsilon} \, F^A\,,
        \qquad%
    \delta F^A = i \, \bar{\epsilon} \, \p_{\bar{z}} \lambda^A\,.
\ee
The supersymmetry transformations are generated by the supercharge operator
$\CQ = \p_{\bar{\theta}} - \bar{\theta} \, \p_{\bar{z}}$ acting on an operator $\CO$ via $\delta_{\bar{\epsilon}} \CO = \, \bar{\epsilon} \, \CQ \, \CO$\,.
In curved spacetime with arbitrary background metric field $g^{}_\text{MN} (X)$\,, Kalb-Ramond field $B^{}_\text{MN} (X)$\,, and YM gauge field $V^{AB}_\text{M} (X)$ in the adjoint representation, the sigma model in the superspace formalism is
\be \label{eq:ssa}
    S = \frac{1}{4\pi} \int \dd^2 z \, d\bar{\theta} \, \Bigl[ \p^{}_z Y^\text{M} \, D^{}_{\bar{\theta}} Y^\text{N} \, \CE^{}_\text{MN} \bigl( Y \bigr) - \Lambda^A \,\nabla^{}_{\!\bar{\theta}} \Lambda^A \Bigr]\,,
\ee
where
\be
    \nabla_{\!\bar{\theta}} \Lambda^A = D_{\bar{\theta}} \Lambda^A - i \, D_{\bar{\theta}} Y^\text{M} \, V_\text{M}^{AB} (Y) \, \Lambda^B\,,
        \qquad%
    \CE^{}_\text{MN} (Y) = g^{}_\text{MN} (Y) + B^{}_\text{MN} (Y)\,. 
\ee
We further introduce the vielbein field $E^{}_\text{M}{}^a$ satisfying $g^{}_\text{MN} = E^{}_\text{M}{}^a \, E^{}_\text{N}{}^b \, \eta^{}_{ab}$\,, with $a = 0\,,\,\cdots,\,9$ the frame index. We define the spin connection $\omega^{}_\text{M}{}^{ab}$ and the three-form field strength $H_\text{MNL}$ to be, respectively,
\begin{subequations}
\begin{align}
    \omega^{}_{\text{M}}{}^{ab} & = 2 \, E^{\text{N}[a} \, \p^{}_\text{[M} E^{}_{\text{N]}}{}^{b]} 
    - E^{\text{N}a} \, E^{\text{L}b} \, E^{}_\text{M}{}^c \, \p^{}_{[\text{N}} E^{}_{\text{L}]c}\,, \\[4pt]
    H_{\text{MNL}} & = \p_\text{M} B_{\text{NL}} + \p_\text{N} B_{\text{LM}} + \p_\text{L} B_{\text{MN}}\,.
\end{align}
\end{subequations}
Here, $\mathcal{O}_{[\text{M}\text{N}]} = \frac{1}{2} ( \mathcal{O}_{\text{M}\text{N}} - \mathcal{O}_{\text{N}\text{M}})$\,. 
In components, the heterotic sigma model~\eqref{eq:ssa} gives
\begin{align} \label{eq:scurv}
\begin{split}
    S = \frac{1}{4\pi} \int \dd^2 z \, \biggl\{ \p_z X^\text{M} \, \p_{\bar{z}} X^\text{N} \, \Bigl[ g^{}_\text{MN} (X) &+ B^{}_\text{MN} (X) \Bigr] + \psi^{}_a \nabla_{\!z} \psi^a \\[4pt]
    + \tr \bigl( \lambda \, \nabla_{\!\bar{z}} \lambda \bigr) & + \tfrac{i}{2} \, \tr \Bigl[ \lambda \, \psi^\text{M} \, \psi^\text{N} \, F^{}_\text{MN} (X) \, \lambda \Bigr] \biggr\}\,,
\end{split}
\end{align}
where $\psi^a = \psi^\text{M} \, E^{}_\text{M}{}^a$ and 
\begin{subequations}
\begin{align}
    \nabla^{}_{\!\bar{z}} \lambda &= \p^{}_{\bar{z}} \lambda - i \, V^{}_{\bar{z}} \, \lambda\,, 
        &%
    F^{}_\text{MN} &= \p^{}_\text{M} V^{}_\text{N} - \p^{}_\text{N} V^{}_\text{M} - i \, \bigl[ V^{}_\text{M}\,, \, V^{}_\text{N} \bigr]\,, \\[4pt]
    \nabla^{}_{\!z} \psi^a &= \p^{}_z \psi^a + {\Omega}^{}_{z}{}^{ab} \, \psi^{}_b\,,
        &%
    {\Omega}^{}_\text{M}{}^{ab} &= \omega^{}_\text{M}{}^{ab} + \tfrac{1}{2} \, H^{}_{\text{NLM}} \, E^{\text{N}a} \, E^{\text{L}b}\,. \label{eq:op}
\end{align}
\end{subequations}
We have defined the pullbacks
$\Omega^{}_z{}^{ab} \equiv \p^{}_z X^\text{M} \, \Omega^{}_\text{M}{}^{ab}$ and $V^{AB}_{\bar{z}} \equiv \partial^{}_{\bar{z}} X^\text{M} \, V^{AB}_\text{M}$\,. There are both YM gauge and gravitational anomalies in the path integral for the sigma model~\eqref{eq:scurv}, which we examine below. 

\paragraph{\emph{Yang-Mills gauge anomalies.}}
In the action~\eqref{eq:scurv}, $V_\text{M}$ is the SO(32) or $\text{E}_8 \times \text{E}_8$ gauge potential. However, our analysis here is insensitive to the specific gauge group. Classically, this action is invariant under the following target space gauge transformation with parameters $\xi^{AB}$ in the adjoint representation:
\begin{equation} \label{eq:gtrnsf}
    \delta_\xi V^{AB}_\text{M} = \nabla^{}_{\!\text{M}} \xi^{AB} \equiv \p^{}_\text{M} \xi^{AB} + i \, \bigl[ \xi\,,\,V^{}_\text{M} \bigr]^{AB}\,,
        \qquad%
    \delta_\xi \lambda^{A} = i \, \xi^{AB} \, \lambda^B\,.
\end{equation}
Here,  $\nabla_{\!\text{M}}$ is the covariant derivative in the target space. 
This classical gauge symmetry acts only on the left-moving fermions in the 2D worldsheet theory and is therefore anomalous when the path integral is concerned. The coupling to the pullback 
$V^{AB}_{\bar{z}}$
of the gauge potential in the action~\eqref{eq:scurv} takes the form,
\begin{equation} \label{eq:inta}
	S_\text{int} = \frac{1}{4\pi} \int \dd^2 z \, \tr \bigl( J_z \, V_{\bar{z}} \bigr)\,,  
\end{equation}
where $J^{AB}_z = i \, \lambda^A (z) \, \lambda^B (z)$ is the current associated with the left-moving fermions. From the operator product expansion (OPE),
$\lambda^A (z) \, \lambda^B (0) \sim \delta^{AB} \, z^{-1}$\,,
we find that, 
\begin{equation}
	J^{AC}_z (z_1) \, J^{BD}_z (z_2) \sim \frac{\delta^{AB} \, \delta^{CD} - \delta^{AD} \, \delta^{BC}}{z_{12}^2} + \cdots\,,   
\end{equation}
where $z_{12} = z_1 - z_2$\,. We consider the path integral
\be \label{eq:pi}
    \CZ = \int \mathscr{D} \bigl[X^\text{M}, \, \lambda^A, \, \psi^a \bigr] \, e^{-S}.
\ee
At the second order of the gauge potential in the path integral, the gauge symmetry is already  anomalous, with 
\begin{equation} \label{eq:dxzzot}
	\delta_\xi \mathcal{Z} = - \frac{1}{8\pi^2} \int \dd^2 z_1 \, \dd^2 z_2 \, \frac{1}{z_{12}^2} \, \tr \Bigl( V_{\bar{z}_1} \nabla_{\!\bar{z}_2} \xi \Bigr). 
\end{equation}
Integrating by parts and using
$\partial_{\bar{z}} ( z^{-2} ) = - 2 \pi \, \partial_z \delta^{(2)} \bigl( z, \bar{z} \bigr)$\,,
we find
\begin{equation} \label{eq:dxz}
	\delta_\xi \mathcal{Z} \sim \frac{1}{4\pi} \int \dd^2 z \, \tr \Bigl( \partial_z \xi \, V_{\bar{z}}\Bigr). 
\end{equation}
In order to compensate this gauge anomaly at the lowest-order quantum corrections, we modify the original action~\eqref{eq:scurv} to be
\begin{align} \label{eq:effaction}
\begin{split}
    S & = \frac{1}{4\pi} \int \dd^2 z \, \Bigl[ \tfrac{2}{\alpha'} \, \p_z X^\text{M} \, \p_{\bar{z}} X^\text{N} \, G^{}_\text{MN} + \psi^{}_a \nabla_{\!z} \psi^a + \tr \bigl( \lambda \, \nabla_{\!\bar{z}} \lambda \bigr) + \tfrac{i}{2} \, \tr \bigl( \lambda \, \psi^\text{M} \, \psi^\text{N} \, F^{}_\text{MN} \, \lambda \bigr) \Bigr]\,,
\end{split}
\end{align}
where we have recovered the $\alpha'$ dependence and introduced a generalized metric,
\begin{equation} \label{eq:mm}
    G^{}_\text{MN} = g^{}_\text{MN} + B^{}_\text{MN} + \frac{\alpha'}{4} \, \tr \Bigl( V^{}_\text{M} \, V^{}_\text{N} \Bigr)\,.
\end{equation}
We also require the following extra gauge transformation of the Kalb-Ramond field~\cite{Hull:1985jv}:
\begin{equation} \label{eq:dxib}
    \delta^{}_\xi B^{}_\text{MN} = - \frac{\alpha'}{4} \, \tr \Bigl[ \xi \, \bigl( \p_\text{M} V_\text{N} - \p_\text{N} V_\text{M} \bigr) \Bigr]\,.
\end{equation}
The gauge-covariant curvature is given in~\cite{Bergshoeff:1995cg} and takes the form of $H^{(3)} = \dd B^{(2)} + \omega^{(3)}$\,, with $\omega^{(3)}$ the 
Chern-Simons three-form constructed from the YM field $V_\text{M}$\,. 
Note that the gauge transformations of the form~\eqref{eq:dxib} close on $B_{\rm MN}$: up to the U(1) gauge transformation $\delta_\zeta B_\text{MN} = \p_{\text{[M}} \zeta_{\text{N]}}$\,, we have
\be
    \bigl( \delta_\xi \delta_{\xi'} - \delta_{\xi'} \delta_{\xi} \bigr) B_\text{MN} 
    = \delta_\Xi B_\text{MN}\,,  
\ee
where $\Xi = i \, ( \xi' \, \xi - \xi \, \xi')$\,.
Even though the modified sigma mode itself is \emph{not} gauge invariant, this modification guarantees that the path integral of the action~\eqref{eq:effaction} is gauge invariant at the lowest order of the quantum corrections.

\paragraph{\emph{Gravitational anomalies.}}
The calculation for gravitational anomalies is analogous to the one for the YM gauge anomalies, except that now essentially the left-moving fermion $\lambda^{AB}$ (and YM gauge potential $V^{AB}$) is replaced with the right-moving fermion $\psi^a$ (and spin connection ${\Omega}_\text{M}{}^{ab}$). Under the Lorentz transformation parametrized by $\ell^{ab}$\,, the path integral~\eqref{eq:pi} gives
\be
    \delta^{}_\ell \CZ \sim \frac{\alpha'}{8\pi} \int \dd^2 z \, \p^{}_{\bar{z}} \ell^{ab} \, \Omega^{}_{zab}\,.
\ee
The one-loop anomaly-free path integral (with respect to both YM gauge and gravitational anomalies) is then associated with same action~\eqref{eq:effaction}, but now with\,\footnote{This generalized metric containing the torsionful spin-connection fields also occurs in a recent derivation of the heterotic T-duality rules using heterotic supergravity with $R^2$ corrections~\cite{Elgood:2020xwu}.}
\begin{equation} \label{eq:mmvvoo}
    G^{}_\text{MN} = g^{}_\text{MN} + B^{}_\text{MN} + \frac{\alpha'}{4} \biggl[ \tr \Bigl( V^{}_\text{M} \, V^{}_\text{N} \Bigr) + \Omega^{}_\text{M}{}^{ab} \, \Omega^{}_{\text{N}ab} \biggr].
\end{equation}
The Kalb-Ramond field also acquires the extra Lorentz transformation~\cite{Hull:1985jv},
\be
   \delta^{}_\ell B^{}_\text{MN} = - \frac{\alpha'}{4} \, \ell^{}_{ab} \, \Bigl( \p^{}_\text{M} {\Omega}^{}_\text{N}{}^{ab} - \p^{}_\text{N} {\Omega}^{}_\text{M}{}^{ab} \Bigr)\,,
\ee
which is analogous to the gauge transformation~\eqref{eq:dxib}. We have thus achieved a modified action whose associated path integral is free of both YM gauge and gravitational anomaly with respect to the lowest-order  quantum corrections.

\subsection{T-Duality Transformations} \label{sec:tdt}

In this section, we show that the T-duality rules found in~\cite{Bergshoeff:1995cg} by using the low-energy effective supergravity in heterotic string theory can be derived using the gauge-invariant path integral that we have discussed in the previous subsection. Ignoring the dilaton term, for which we will come back at the end of this subsection, we take the action~\eqref{eq:effaction} in conformal gauge as our starting point.  
In order to perform a T-duality transformation, we introduce a Killing vector $k^\text{M}$\,, which we assume to satisfy the condition $G^{}_\text{MN} \, k^\text{M} \, k^\text{N} \neq 0$\,. Note that the Killing vector is defined with respect to the generalized metric $G^{}_\text{MN}$ rather than the target space metric $g^{}_\text{MN}$\,. It is convenient to define the adapted coordinates $X^\text{M} = (X^\mu_{}\,, \, x)$\,, where $\mu$ is an (8+1)D spacetime index, such that $\nabla_{\!x} = k^\text{M} \, \nabla_\text{\!M}$ with $\nabla$ the covariant derivative with respect to the Yang-Mills gauge field $V_\text{M}^{AB}$. We call $x$ an isometry direction in this modified sense.  

In the following, we focus on the NSNS sector. In this regard, it is sufficient for us to only consider the ($X^\text{M}$, $\lambda$) sector while setting $\psi^a$ to zero. 
Consider the path integral associated with the action~\eqref{eq:effaction} with the generalized metric~\eqref{eq:mmvvoo} (but ignore the dependence on $\psi$), 
\be \label{eq:zmod}
    \CZ = \int \mathscr{D} \bigl[ X^\text{M}, \, \lambda^A \bigr] \, e^{-S}\,,
\ee
which we have shown to be free of YM gauge and gravitational anomaly. 
In the presence of the isometry, the modified action can be seen as descending from the `parent' action~\cite{Buscher:1987sk},
\begin{align} \label{eq:parentaction}
\begin{split}
    S_{\rm parent} = \frac{1}{4\pi} \int \dd^2 z \, \biggl\{ & \partial_z X^{\mu} \, \partial_{\bar{z}} X^{\nu} \, G_{\mu\nu} + \chi^{}_+ \, \chi^{}_- \, G_{xx} 
    + \chi^{}_+ \, \partial_{\bar{z}} X^{\mu} \, G_{x \mu} + \partial_z X^{\mu} \, \chi^{}_- \, G_{\mu x} \\[4pt]
    & + \tr \Bigl[ \lambda \bigl( \partial_{\bar{z}} \lambda - i \, V_\mu \, \partial_{\bar{z}} X^\mu \lambda - i \, V_x \, \chi^{}_- \, \lambda \bigr) \Bigr] 
    + \tilde{x} \, \bigl( \partial_z \chi^{}_- -\partial_{\bar{z}} \chi^{}_+ \bigr) \biggr\}\,.
\end{split}
\end{align}
We have introduced the Lagrange multiplier $\tilde{x}$ imposing the Bianchi identity $\partial^{}_z \chi^{}_- = \partial^{}_{\bar{z}} \chi^{}_+$\,. The field $\tilde{x}$ will play the role of the coordinate dual to $x$. Now, the path integral~\eqref{eq:zmod} is written equivalently as
\be \label{eq:zmt}
    \CZ = \int \mathscr{D}\! \left[ X^\mu, \, \tilde{x}\,, \, \lambda^A, \, \chi^{}_+\,, \, \chi^{}_- \right] \, e^{-S_\text{parent}}\,.
\ee
Integrating out $\tilde{x}$\,, the Bianchi identity can be solved locally by $\chi^{}_+ = \partial^{}_z x$ and $\chi^{}_- = \partial^{}_{\bar{z}} x$\,, such that the original action~\eqref{eq:effaction} is reproduced. To derive the dual sigma model, we instead integrate out $\chi^{}_\pm$ in the parent action~\eqref{eq:parentaction}, 
which gives rise to the dual action
\begin{align} \label{eq:dualaction}
    S_\text{dual} = \! \frac{1}{4\pi} \int \! \dd^2 z \, \Bigl( \partial_z \tilde{X}^\text{M} \, \partial_{\bar{z}} \tilde{X}^\text{N} \, \tilde{G}_\text{MN} + \lambda^A \tilde{D}_{\bar{z}} \lambda^A \Bigr)\,,
\end{align}
where $\tilde{X}^\text{M} = (X^\mu\,, \, \tilde{x})$ and
$\tilde{D}_{\bar{z}} \lambda = \partial_{\bar{z}} \lambda - i \, \tilde{V}_\text{M} \, \partial_{\bar{z}} \tilde{X}^\text{M} \, \lambda$\,.
The heterotic Buscher rules are
\begin{subequations} \label{eq:hetbuscher}
\begin{align}
    \tilde{G}_{xx} &= \frac{1}{G_{xx}}\,, 
        &%
    \tilde{G}_{\mu x} &= - \frac{G_{\mu x}}{G_{xx}}\,,
        &
    \tilde{V}_{x} &= - \frac{V_{x}}{G_{xx}}\,, \\[4pt]
    \tilde{G}_{\mu\nu} &= G_{\mu\nu} - \frac{G_{\mu x} \, G_{x \nu}}{G_{xx}}\,,
        &%
    \tilde{G}_{x \mu} &= \frac{G_{x \mu}}{G_{xx}}\,,
        &
    \tilde{V}_{\mu} &= V_{\mu} - \frac{V_x \, G_{x \mu}}{G_{xx}}\,.
\end{align}
\end{subequations}
Alternatively, one may gauge the isometry as in~\cite{Rocek:1991ps} to obtain the same Buscher rules above. 

The T-duality transformation also acts nontrivially on the dilaton field, which we have not yet discussed. The inclusion of the dilaton requires undoing the conformal gauge and considering a curved worldsheet. We thereby introduce the worldsheet zweibein fields $e_\alpha{}^1$ and $e_\alpha{}^2$\,, with the `curved' index $\alpha = 1\,,\,2$ on the worldsheet. We also introduce the worldsheet coordinates $\sigma^\alpha$\,. The covariantization is then implemented via the following replacements:
\begin{align} \label{eq:cov}
    2 \, \p_z & \rightarrow \CD \equiv \frac{i}{e} \, \epsilon^{\alpha\beta} \, \bar{e}_\alpha \, \p_\beta\,, 
        &%
    2 \, \p_{\bar{z}} & \rightarrow \bar{\CD} \equiv \frac{i}{e} \, \epsilon^{\alpha\beta} \, e_\alpha \, \p_\beta\,, 
        &%
    \dd^2 z & \rightarrow 2 \, \dd^2 \sigma \, e\,,
\end{align}
where $e_\alpha = e_\alpha{}^1 + i \, e_\alpha{}^2$\,, $\bar{e}_\alpha = - e_\alpha{}^1 + i \, e_\alpha{}^2$\,, and $e = \det \bigl(e_\alpha{}^a \bigr)$\,. The worldsheet metric is
$h_{\alpha\beta} = e_\alpha{}^1 \, e_\beta{}^1 + e_\alpha{}^2 \, e_\beta{}^2$\,. 
The sigma model~\eqref{eq:effaction} then becomes
\begin{align} \label{eq:effactionm}
\begin{split}
    S & = \frac{1}{8\pi} \int \dd^2 \sigma \, e \, \biggl\{ \, \CD X^\text{M} \, \bar{\CD} X^\text{N} \, G_\text{MN} + 2 \, \tr \Bigl[ \lambda \, \bigl( \bar{\CD} \lambda - i \, V_\text{M} \, \bar{\CD} X^\text{M} \, \lambda \bigr) \Bigr] \biggr\} \\[4pt]
    & \quad + \frac{1}{4\pi} \int \dd^2 \sigma \, \sqrt{h} \, R (h) \, \Phi\,.
\end{split}
\end{align}
Now, T-dualizing along the isometry $x$ leads to the same Buscher rules as in Eq.~\eqref{eq:hetbuscher}. In addition, 
the analog of the $\chi_+ \, \chi_- \, G_{xx}$ term in Eq.~\eqref{eq:parentaction} contributes to a shift of the dilaton after integrating out $\chi_\pm$ in the path integral. This shift can be evaluated by using the heat kernel method as in~\cite{Buscher:1987qj} and yields the standard Buscher rule
$\tilde{\Phi} = \Phi - \frac{1}{2} \ln G_{xx}$\,.
Note that the dual action is of the same form as the original action~\eqref{eq:effactionm}. 

The above worldsheet method provides a much simpler derivation that recovers the heterotic Buscher rules previously found by dimensionally reducing heterotic supergravity~\cite{Bergshoeff:1995cg}.

\subsection{Heterotic Non-Relativistic String Theory from DLCQ} \label{sec:nonhstdlcq}

\begin{figure}[t!]
\centering
\begin{tikzpicture}
    \begin{scope}[scale=.8]

        \path[every node/.style=draw, rounded corners=1, line width=1.5pt, minimum width=140pt, minimum height=20pt, font = \small]   
                (3-.5,1.5) node {DLCQ of IIA/B superstring}
                (11+.5,1.5) node {IIB/A non-relativistic string}
                (3-.5,3.7) node {IIA/B superstring};

            \path[every node/.style=draw, rounded corners=1, line width=1.5pt, minimum width=140pt, minimum height=20pt, font = \small]
                (11+.5,3.7) [align=center] node {IIB/A superstring};

        \path   (7,3.2+0.75) [font=\footnotesize] node {\scalebox{0.9}{\emph{T-dual}}}
                (2.6,2.6) [font=\footnotesize] node {\emph{infinite boost limit}}
                (11.6,2.6) [font=\footnotesize] node {\emph{BPS decoupling limit}}
                (7,2.1) [font=\footnotesize] node {\scalebox{0.9}{\emph{long. spatial}}}
                (7,1.5) [font=\footnotesize] node {\scalebox{0.9}{\emph{T-dual}}}
                (7,0.85) [font=\footnotesize] node {\scalebox{0.9}{$g^{}_\text{MN} \, k^\text{M} \, k^\text{N} = 0$}}
                ;

        \draw[{<[length=1.3mm]}-{>[length=1.3mm]}] (5.6,3.5+0.2) -- (8.4,3.5+0.2);
        \draw[{<[length=1.3mm]}-{[length=1.3mm]}] (5.6,3.5+0.3-2) -- (8.4,3.5+0.3-2);
        \draw[{[length=1.3mm]}-{>[length=1.3mm]}] (5.6,3.5-0.3-2) -- (8.4,3.5-0.3-2);
       
        \draw [{<[length=1.3mm]}-{[length=1.3mm]}] (4.75-.4,2) -- (4.75-.4,3.2);
        \draw [{<[length=1.3mm]}-{[length=1.3mm]}] (9+.6,2) -- (9+.6,3.2);
        \draw (0,0) node {\phantom{c}};
\end{scope}
\end{tikzpicture}
\begin{tikzpicture}
    \begin{scope}[scale=.8]

        \path[every node/.style=draw, rounded corners=1, line width=1.5pt, minimum width=140pt, minimum height=20pt, font = \small]   
                (3-0.5,1.5) node {DLCQ of heterotic superstring}
                (11+0.5,1.5) node {\textbf{heterotic non-rel. string?}};
            
        \path   (7.05,1.75) [font=\footnotesize] node {\scalebox{0.9}{\emph{T-dual}}}
                (7.05,1.1) [font=\footnotesize] node {\scalebox{0.9}{$G^{}_\text{MN} \, k^\text{M} \, k^\text{N} = 0$}}
                ;

        \draw[{[length=1.3mm]}-{>[length=1.3mm]}] (5.6,3.5-2) -- (8.4,3.5-2);
\end{scope}
\end{tikzpicture}
\caption{Heterotic non-relativistic string sigma models can be built by performing a T-duality transformation in the DLCQ of heterotic string theory along a lightlike isometry direction defined with respect to the generalized metric $G^{}_\text{MN}$\,. This is analogous to the connection between the DLCQ of type II superstring theory and its associated non-relativistic string theory, where in the latter case the lightlike isometry direction is defined with respect to the ordinary metric $g^{}_\text{MN}$\,. The BPS decoupling limit relating type II relativistic to non-relativistic superstring theory will be discussed later in Section~\ref{sec:ccdbps}.}
\label{fig:rm0}
\end{figure}
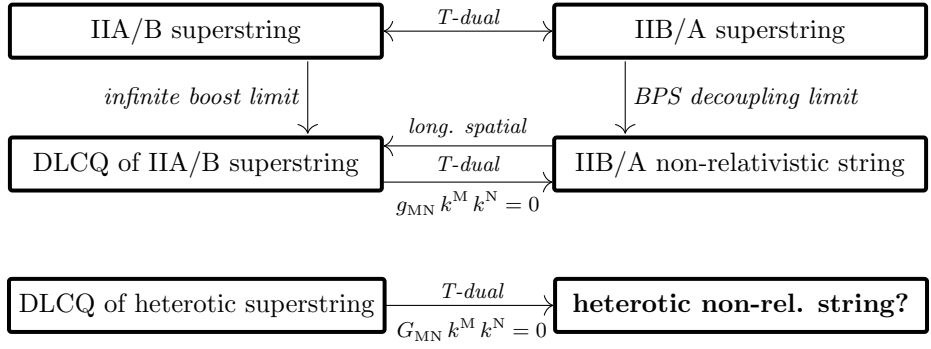

In Section~\ref{sec:tdt}, we performed a T-duality transformation along an isometry direction associated with a Killing vector $k^\text{M}$ satisfying $G_\text{MN} \, k^\text{M} \, k^\text{N} \neq 0$\,. The complementary case satisfying 
\be \label{eq:gxxnull}
    G_{xx} = G_\text{MN} \, k^\text{M} \, k^\text{N} = 0
\ee
implies that $x$ is a lightlike isometry with respect to the generalized metric $G_{xx}$ and thus leads to the DLCQ of heterotic string theory. This is in contrast to type II superstring theory in the DLCQ, where it is the target space metric $g^{}_\text{MN}$ that develops a lightlike isometry with $g_{xx} = 0$\,. In order to understand this singular choice of $G_{xx} = 0$\,, we first briefly review the type II analog and its dual description of non-relativistic string theory. See Figure~\ref{fig:rm0} for an illustration for this distinction between the type II and heterotic cases. 

\subsubsection{Review of Gomis-Ooguri Sigma Models} \label{sec:rgosm}

We now consider type II superstring theory in the DLCQ, where there is a lightlike compactification in the target space. The DLCQ of quantum field theories sometimes leads to pathological behaviors at the quantum level~\cite{Hellerman:1997yu, Chapman:2020vtn}. However, in string theory, the DLCQ can be defined from first principles using its dual description, where the lightlike circle is mapped to a conventional spacelike compactification, while the target space becomes non-Lorentzian~\cite{Bergshoeff:2018yvt, Bergshoeff:2019pij}.\,\footnote{See also discussions using the Hamiltonian formulation~\cite{Kluson:2018vfd} as well as from the open string~\cite{Gomis:2020izd, Ebert:2021mfu} and null reduction~\cite{Harmark:2017rpg, Kluson:2018egd, Harmark:2018cdl, Harmark:2019upf} perspective.} Such a dual description is referred to as non-relativistic string theory~\cite{Klebanov:2000pp, Gomis:2000bd, Danielsson:2000gi} since the closed string spectrum enjoys a (string) Galilean invariance. The second quantization of type IIB non-relativistic superstring theory is matrix string theory~\cite{Motl:1997th, Dijkgraaf:1997vv}, which is dual to BFSS matrix quantum mechanics. 

We start with the bosonic sector of the type II string sigma model, 
\be
    S^{}_\text{II} = \frac{1}{4\pi} \int \dd^2 z \, \p^{}_z X^\text{M} \, \p^{}_{\bar{z}} X^\text{N} \bigl( g^{}_\text{MN} + B^{}_\text{MN} \bigr).
\ee
When there is a lightlike Killing vector $\ell^\text{M}$ satisfying $g^{}_\text{MN} \, \ell^\text{M} \, \ell^\text{N} = 0$\,, we define the adapted coordinates $X^\text{M} = (X^\mu, \, x)$ with $\p_x = \ell^\text{M} \, \p_\text{M}$\,. In the DLCQ, we compactify the lightlike isometry direction $x$\,. Following Section~\ref{sec:tdt}, we introduce the parent action, 
\begin{align} \label{eq:parentactionnrst}
\begin{split}
    S_{\rm parent} &= \frac{1}{4\pi} \int \dd^2 z \, \Bigl[ \partial_z X^{\mu} \, \partial_{\bar{z}} X^{\nu} \, \bigl( g^{}_{\mu\nu} + B^{}_{\mu\nu} \bigr)  
    + \, \chi^{}_+ \, \partial_{\bar{z}} X^{\mu} \, \bigl( g^{}_{x \mu} + B^{}_{x \mu} \bigr) \\[4pt]
    & \hspace{5cm} + \partial_z X^{\mu} \, \chi^{}_- \, \bigl( g^{}_{\mu x} + B^{}_{\mu x} \bigr) + \tilde{x} \, \bigl( \partial_z \chi^{}_- - \partial^{}_{\bar{z}} \chi^{}_+ \bigr) \Bigr]\,,
\end{split}
\end{align}
such that integrating out the Lagrange multiplier $\tilde{x}$ gives back the original action describing the type II string in the DLCQ.
Unlike Eq.~\eqref{eq:parentaction}, this parent action does not contain any $\chi_+ \, \chi_-$ term.  In order to pass on to the T-dual frame, we integrate by parts and
rewrite the parent action~\eqref{eq:parentactionnrst} in the dual form,
\be \label{eq:dualnrst}
    S^{}_\text{NRST} = \frac{1}{4\pi} \int \dd^2 z \, \Bigl( \p^{}_z \tilde{X}^\text{M} \, \p^{}_{\bar{z}} \tilde{X}^\text{N} \, \CE^{}_{\text{M}\text{N}} + \chi^{}_+ \, \p^{}_{\bar{z}} \tilde{X}^\text{M} \, \tau^{}_\text{M}{}^+ + \chi^{}_- \, \p^{}_{z} \tilde{X}^\text{M} \, \tau^{}_\text{M}{}^- \Bigr)\,,
\ee
with $\tilde{X}^\text{M} = (X^\mu, \tilde{x}\,)$ and
\begin{subequations}
\begin{align}
    \CE^{}_{\mu\nu} &= g^{}_{\mu\nu} + B^{}_{\mu\nu}\,,
        &%
    \tau^{}_{\mu}{}^+ &= g^{}_{x\mu} + B^{}_{x\mu}\,,
        &%
    \tau^{}_{\mu}{}^- &= g^{}_{x\mu} - B^{}_{x\mu}\,, \\[4pt]
    \CE^{}_{{\tilde{x}}\text{M}} &= \CE^{}_{\text{M}{\tilde{x}}} = 0\,,
        &%
    \tau^{}_{\tilde{x}}{}^+ &= 1\,,
        &%
    \tau^{}_{\tilde{x}}{}^- &= -1\,.
\end{align}
\end{subequations}
Note that $\chi_\pm$ now act as Lagrange multipliers imposing $\p^{}_{\bar{z}} \tilde{X}^\text{M} \, \tau^{}_\text{M}{}^+ = \p^{}_{z} \tilde{X}^\text{M} \, \tau^{}_\text{M}{}^- = 0$\,. 
The dual sigma model~\eqref{eq:dualnrst} is already in the form of the curved-background generalization~\cite{Bergshoeff:2018yvt} of the Gomis-Ooguri action~\cite{Gomis:2000bd} describing non-relativistic string theory. We will return to the Buscher rule for the dilaton after illustrating the dual geometry.

\begin{figure}[bt!]
    \centering
    \includegraphics[scale=0.6]{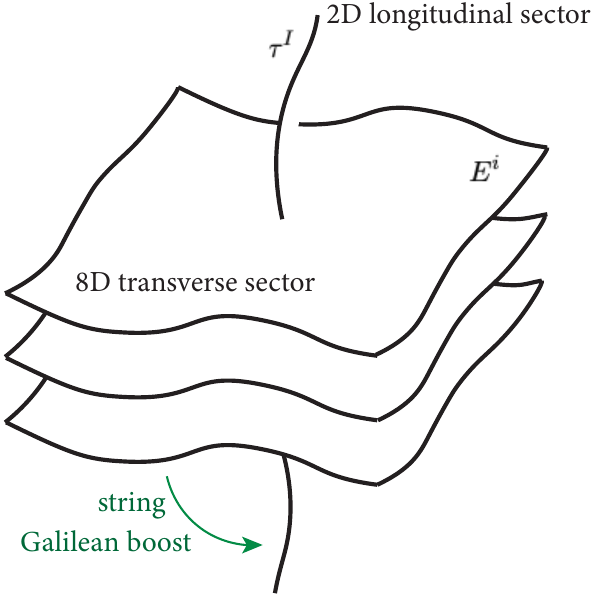}
    \caption{This is a sketch of string Newton-Cartan geometry. T-dualizing along a spatial circle within the 2D longitudinal sector gives back the DLCQ of the 10D Lorentzian target space geometry in superstring theory.  
    This non-Lorentzian string Newton-Cartan geometry is characterized by a string Galilean boost that maps from the transverse to longitudinal sector. This boost generalizes the Galilean boost to include a Galilei-like rotation from a transverse to longitudinal spatial direction.}
    \label{fig:snc}
\end{figure}

Generically, $\CE^{}_\text{MN}$ and $\tau^{}_\text{M}{}^\pm$ are background fields that are arbitrary functions of the worldsheet fields $\tilde{X}^\text{M}$\,. 
Upon appropriate redefinitions of the Lagrange multipliers $\chi^{}_\pm$\,, the dual action~\eqref{eq:dualnrst} describes non-relativistic string theory in general background fields, with
\be \label{eq:eeeb}
    \CE^{}_\text{MN} = E^{}_\text{M}{}^{i} \,  E^{}_\text{N}{}^{i} + b^{}_\text{MN}\,,
        \qquad%
    i = 2\,, \, \cdots, \, 9\,,
\ee
where $b_\text{MN} = \CE_{[\text{MN}]}$ is the Kalb-Ramond two-form potential. We further define
\be \label{eq:lv}
    \tau^{}_\text{M}{}^0 = \tfrac{1}{2} \, \bigl( \tau^{}_\text{M}{}^+ + \tau^{}_\text{M}{}^- \bigr)\,,
        \qquad%
    \tau^{}_\text{M}{}^1 = \tfrac{1}{2} \, \bigl( \tau^{}_\text{M}{}^+ - \tau^{}_\text{M}{}^- \bigr)\,.
\ee
The target space geometry is now equipped with a codimension-two foliation structure\,\footnote{Strictly speaking, the target space only has a foliation if there are extra torsional constraints imposed on the longitudinal vielbein field $\tau^{}_\text{M}{}^I$, such that it is integrable. In general, this is a distribution.} encoded by the longitudinal vielbein field $\tau^{}_\text{M}{}^I$, $I = 0\,,\,1$\,, and the transverse vielbein field $E^{}_\text{M}{}^i$\,. The 8D transverse sector is related to the 2D longitudinal sector via a stringy generalization of the Galilean boosts, which act on the vielbein fields as
\begin{subequations} \label{eq:sgb}
\be 
    \delta^{}_\text{G} \tau^{}_\text{M}{}^I = 0\,,
        \qquad%
    \delta^{}_\text{G} E^{}_\text{M}{}^{i} = \Lambda^{}_I{}^{i} \, \tau^{}_\text{M}{}^I\,.
\ee
Moreover, the string Galilean boosts also act non-trivially on the $B$-field $b^{}_\text{MN}$ and the Lagrange multipliers $\chi^{}_\pm$\,, with
\begin{align} 
    \delta^{}_\text{G} b^{}_\text{MN} & = - 2 \, \Lambda^{}_I{}^i \, E^{}_{[\text{M}}{}^i \, \tau^{}_{\text{N}]}{}^J \, \epsilon^{I}{}^{}_{J}\,, 
        &%
    \begin{array}{c}
        \delta \chi^{}_+ = - \Lambda^{}_+{}^i \, \p^{}_z \tilde{X}^\text{M} E_\text{M}{}^i\,, \\[10pt]
        \delta \chi^{}_- = - \Lambda^{}_-{}^i \, \p^{}_{\bar{z}} \tilde{X}^\text{M} E_\text{M}{}^i\,,
    \end{array}
\end{align}
\end{subequations}
with $\Lambda^{}_\pm{}^i = \Lambda^{}_0{}^i \pm \Lambda^{}_1{}^i$\,. 
The vielbein fields define the string Newton-Cartan geometry in the target space~\cite{Andringa:2012uz, Bergshoeff:2018yvt, Bergshoeff:2019pij, Bidussi:2021ujm}. There is \emph{no} 10D metric description, which implies that the string Newton-Cartan geometry is non-Lorentzian. See Figure~\ref{fig:snc} for an illustration. Even though the target space is non-Lorentzian, the worldsheet theory of non-relativistic string theory is Lorentzian and standard conformal field theoretical techniques still apply~\cite{Gomis:2000bd, Danielsson:2000mu}. Quantum mechanically, the sigma models are only self-contained under a sufficiently strong symmetry principle, in which case a geometric constraint on the longitudinal vielbein field $\tau^{}_\text{M}{}^I$ is imposed~\cite{Yan:2021lbe}. Such a geometric constraint is also essential for preserving target space supersymmetry~\cite{Bergshoeff:2021tfn, Bergshoeff:2024nin}. 
See Section~\ref{sec:ccdbps} for further discussions. 

Note that in this dual frame described by non-relativistic string theory, the lightlike circle in the original theory is now mapped to a longitudinal spatial circle. As a unitary and ultra-violet complete string theory~\cite{Gomis:2000bd}, non-relativistic string theory provides a first-principles definition for strings in the DLCQ via a longitudinal spatial T-duality transformation~\cite{Bergshoeff:2018yvt}. 

Finally, we comment on the Buscher rule for the dilaton following closely~\cite{Bergshoeff:2018yvt}. Following the logic that we use non-relativistic string theory as a definition of the DLCQ of type II string theory, it is easier to T-dualize non-relativistic string theory along the longitudinal spatial compactification. By redefining the Lagrange multipliers $\chi^{}_\pm$ in Eq.~\eqref{eq:dualnrst}, we switch to an equivalent description where Eq.~\eqref{eq:eeeb} is modified to be
\be
    \CE^{}_\text{MN} = H^{}_\text{MN} + b'_\text{MN}\,,
        \qquad%
    H^{}_\text{MN} = E^{}_\text{M}{}^i \, E^{}_\text{N}{}^i + \Bigl( \tau^{}_\text{M}{}^I \, m^{}_\text{N}{}^J + \tau^{}_\text{N}{}^I \, m^{}_\text{M}{}^J \Bigr) \, \eta^{}_{IJ}\,, 
\ee
such that $H^{}_{xx}$ is no longer zero. Here, we have introduced a new field $m^{}_\text{M}{}^I$ associated with the stringy generalization of the particle number generator in the gauging of the string Bargmann algebra~\cite{Andringa:2012uz, Bergshoeff:2019pij}. This extra field is associated with the conservation of string winding number. Now, we return to the non-relativistic string sigma model~\eqref{eq:dualnrst}. Dropping the tildes and undoing the conformal gauge as in Eq.~\eqref{eq:effactionm}, we write
\begin{align} \label{eq:dualnrstudc}
\begin{split}
    S^{}_\text{NRST} &= \frac{1}{8\pi} \int \dd^2 \sigma \, e \Bigl( \CD X^\text{M} \, \bar{\CD} X^\text{N} \, \CE^{}_{\text{M}\text{N}} + 2 \, \chi^{}_+ \, \bar{\CD} X^\text{M} \, \tau^{}_\text{M}{}^+ + 2 \, \chi^{}_- \, \CD X^\text{M} \, \tau^{}_\text{M}{}^- \Bigr) \\[4pt]
    & \quad + \frac{1}{4\pi} \int \dd^2 \sigma \, \sqrt{h} \, R(h) \, \Phi\,.
\end{split}
\end{align}
Now, we introduce a longitudinal spatial isometry $x$ and the associated adapted coordinates $X^\text{M} = (X^\mu,\,x)$\,, such that $\tau^{}_x{}^0 = E^{}_x{}^i = 0$ and $\tau^{}_x{}^1 \neq 0$\,. In the parent action akin to Eq.~\eqref{eq:parentaction}, we replace $\p^{}_\alpha x$ with $v^{}_\alpha$ and introduce a dual coordinate $\tilde{x}$ playing the role of a Lagrange multiplier, which imposes the Bianchi identity $\epsilon^{\alpha\beta} \p^{}_\alpha v^{}_\beta = 0$\,. We now observe that there exists a term containing the factor $v^{}_\alpha \, v^\alpha \, H^{}_{xx}$\,, which is quadratic in $v^{}_\alpha$\,. Integrating out $v^{}_\alpha$ in the path integral then contributes a shift to the dilaton in the same way as discussed at the end of Section~\ref{sec:tdt}, with $\Phi \rightarrow \Phi - \frac{1}{2} \log H_{xx}$\,. In addition, the resulting action now contains a term proportional to $\chi^{}_+ \, \chi^{}_- \, ( \tau^{}_x{}^1 )^2 / H^{}_{xx}$\,. Finally, we also integrate out $\chi^{}_\pm$ in the path integral, which gives rise to the T-dual frame described by the DLCQ of conventional string theory. The term quadratic in $\chi^{}_\pm$ gives rise to a second shift of the dilaton, which combines with the first shift into the final Buscher rule for the dilaton, with $\Phi \rightarrow \Phi - \log |\tau^{}_x{}^1|$\,. Note that the resulting Buscher rules are non-singular in the $H^{}_{xx} \rightarrow 0$ limit.  We refer the readers to~\cite{Bergshoeff:2018yvt} for further details. 

\subsubsection{Heterotic Non-Relativistic String Sigma Models} \label{sec:hnrssm}

\paragraph{\emph{T-dual of the DLCQ.}} Now, we return to the null condition~\eqref{eq:gxxnull} associated with the DLCQ of heterotic string sigma models. Recall that the generalized metric $G^{}_\text{MN}$ in this null condition is the modified one in Eq.~\eqref{eq:mmvvoo}. Rather than a lightlike isometry in the background geometry, in the adapted coordinates we now get
\be \label{eq:nn}
    g^{}_{xx} = - \frac{\alpha'}{4} \, \Bigl[ \tr \bigl( V_x^2 \bigr) + \Omega^+_x{}^{ab} \, \Omega^+_{xab}\Bigr]\,. 
\ee
Under this condition, the parent action~\eqref{eq:parentaction} becomes
\begin{align} \label{eq:parentactiont}
\begin{split}
    S_{\rm parent} = \frac{1}{4\pi} \int \dd^2 z \, \biggl\{ & \partial_z X^{\mu} \, \partial_{\bar{z}} X^{\nu} \, G_{\mu\nu}  
    + \chi^{}_+ \, \partial_{\bar{z}} X^{\mu} \, G_{x \mu} + \partial_z X^{\mu} \, \chi^{}_- \, G_{\mu x} \\[4pt]
    & + \tr \Bigl[ \lambda \bigl( \partial_{\bar{z}} \lambda - i \, V_\mu \, \partial_{\bar{z}} X^\mu \lambda - i \, V_x \, \chi^{}_- \, \lambda \bigr) \Bigr] 
    + \tilde{x} \, \bigl( \partial_z \chi^{}_- - \partial_{\bar{z}} \chi^{}_+ \bigr) \biggr\}\,.
\end{split}
\end{align}
Just like the parent action~\eqref{eq:parentactionnrst} that leads to non-relativistic string theory, the action~\eqref{eq:parentactiont} does not contain any $\chi^{}_+ \, \chi^{}_-$ term. This signals that the T-dual theory should acquire a non-Lorentzian geometric structure in the target space. In contrast, when Eq.~\eqref{eq:nn} is violated, we are back to the parent action~\eqref{eq:parentactionnrst} that does contain a $\chi^{}_+ \, \chi^{}_-$ term and thus the standard heterotic Buscher rules that we have already derived in Eq.~\eqref{eq:hetbuscher}. 

Upon integration by parts, the parent action~\eqref{eq:parentactiont} gives the dual worldsheet theory,
\begin{align} \label{eq:nrhst}
\begin{split}
	S^{}_\text{dual} = \frac{1}{4\pi} \int \dd^2 z \, \biggl\{ & \p^{}_z \tilde{X}^\text{M} \, \p^{}_{\bar{z}} \tilde{X}^\text{N} \, \CE^{}_{\text{M}\text{N}} + \tr \Bigl( \lambda \, \p^{}_{\bar{z}} \lambda - i \, \lambda \, u^{}_{\text{M}} \, \p^{}_{\bar{z}} \tilde{X}^\text{M} \, \lambda \Bigr) \\[4pt]
    & + \chi^{}_+ \, \p^{}_{\bar{z}} \tilde{X}^{\text{M}} \, \tau^{}_{\text{M}}{}^+ + {\chi}^{}_- \, \Bigl[ \p^{}_z \tilde{X}^\text{M} \, {\tau}^{}_{\text{M}}{}^- + i \, \tr \bigl( \lambda \, v^{}_+ \, \lambda \bigr) \Bigr]  \biggr\}\,,
\end{split}
\end{align}
with $\tilde{X}^\text{M} = (X^\mu, \tilde{x})$ and the background fields
\begin{subequations}
\begin{align}
    \CE^{}_{\mu\nu} &= G^{}_{\mu\nu}\,,
        &%
    \tau^{}_\mu{}^+ &= G^{}_{x\mu}\,,
        &%
    \tau^{}_\mu{}^- &= G^{}_{\mu x}\,,
        &%
    u^{}_\mu &= V^{}_\mu\,, \\[4pt]
    \CE^{}_{\tilde{x}\text{M}} &= \CE^{}_{\text{M}\tilde{x}} = 0\,,
        &%
    \tau^{}_{\tilde{x}}{}^+ &= 1\,,
        &%
    \tau^{}_{\tilde{x}}{}^- &= -1\,,
        &%
    u^{}_{\tilde{x}} &= 0\,,
        &%
    v^{}_+ &= - V^{}_x\,.
\end{align}
\end{subequations}
This dual action motivates the heterotic version of non-relativistic string sigma models, which we study in detail below. We will start with the classical analysis without including the $O(\alpha')$ contribution in the generalized metric~\eqref{eq:mmvvoo}. We will then analyze the gauge anomaly in heterotic non-relativistic string sigma models in an intrinsic way, which will allow us to fix the dependence on the YM vector in the quantum action. We will return to the gravitational anomalies later after we present the complete supersymmetric sigma models in Section~\ref{sec:ccdhst}. 

\paragraph{\emph{Classical action.}}
We now focus on the classical string action without the $O(\alpha')$ contribution in the generalized metric~\eqref{eq:mmvvoo}. Dropping the tildes in Eq.~\eqref{eq:nrhst}, we write 
\begin{align} \label{eq:nrhst123}
\begin{split}
	S^{}_\text{hNRST} = \frac{1}{4\pi} \int \dd^2 z \, \biggl\{ & \p^{}_z X^\text{M} \, \p^{}_{\bar{z}} X^\text{N} \, \CE^{}_\text{MN} + \tr \Bigl( \lambda \, \p^{}_{\bar{z}} \lambda - i \, \lambda \, \p^{}_{\bar{z}} X^\text{M} \, u^{}_{\text{M}} \, \lambda \Bigr) \\[4pt]
    & \qquad\,\,\,\, + \chi^{}_+ \, \p^{}_{\bar{z}} X^{\text{M}} \, \tau^{}_{\text{M}}{}^+ + {\chi}^{}_- \, \Bigl[ \p^{}_z X^\text{M} \, {\tau}^{}_{\text{M}}{}^- + i \, \tr \bigl( \lambda \, v^{}_+ \, \lambda \bigr) \Bigr]  \biggr\}\,,
\end{split}
\end{align}
where $\CE^{}_\text{MN} = E^{}_{\text{M}\text{N}} + b^{}_\text{MN}$ and the background fields are functions of $X^\text{M}$\,. It is always possible to redefine the Lagrange multipliers $\chi^{}_\pm$ to bring the background field $E^{}_\text{MN}$ into the form
$E^{}_\text{MN} = E^{}_\text{M}{}^i \, E^{}_\text{N}{}^i$ with $E^{}_\text{M}{}^i$ the longitudinal vielbein field and $i = 2\,,\, \cdots, \, 9$\,.
Now, $E^{}_\text{MN}$ is of rank 8. Together with the longitudinal vielbein fields $\tau^{}_\text{M}{}^\pm$\,, we recover the string Newton-Cartan geometry in the target space, analogous to the type II case in Section~\ref{sec:rgosm}. Upon rescaling $\chi^{}_\pm$\,, the action~\eqref{eq:nrhst123} is invariant under the 2D Lorentz boost transformations $\tau^{}_\text{M}{}^+ \rightarrow \gamma \, \tau^{}_\text{M}{}^+$, $\tau^{}_\text{M}{}^- \rightarrow \tau^{}_\text{M}{}^-/ \gamma$\,, and $v^{}_+ \rightarrow v^{}_+ / \gamma$\,, where $\gamma$ parameterizes the boost.
The YM gauge transformations act on the background YM field $u^{}_\text{M}$ and the worldsheet field $\lambda^A$ as
\begin{subequations} \label{eq:gthsnr}
\be
    \delta^{}_\xi u^{}_\text{M} = \nabla^{}_{\!\text{M}} \xi^{}\,,
        \qquad%
    \delta^{}_\xi \lambda = i \, \xi \, \lambda\,,
\ee
where the gauge covariant derivative has been defined in Eq.~\eqref{eq:gtrnsf}.
Requiring that the action~\eqref{eq:nrhst123} be invariant under the above gauge symmetry, it follows that $v^{}_+$ has to transform non-trivially as
\be
    \delta^{}_\xi v^{}_+ = i \, [\xi\,,\,v^{}_+]\,. 
\ee
\end{subequations}
Apart from the YM gauge transformations, the string Galilean boost~\eqref{eq:sgb} is also modified in the heterotic case, which now additionally acts on the gauge field $u^{}_\text{M}$ as
\be \label{eq:dgum}
    \delta^{}_\text{G} u^{}_\text{M} = - \Lambda^{}_-{}^i \, E^{}_\text{M}{}^i \, v^{}_+\,. 
\ee
This modification is required to compensate  
the boost transformation of $\chi_- \, \tr (\lambda \, v_+ \lambda)$ in the action~\eqref{sec:rgosm}, as $\chi^{}_-$ transforms non-trivially in Eq.~\eqref{eq:sgb}.

In Appendix~\ref{app:ss}, we show how a Stueckelberg symmetry allows one to eliminate the $u_+$ component of $u_\text{M}$\,, leaving us with the usual number of components in the gauge multiplet $(v_+\,,\,u_-\,,\,u_i)$\,. This formulation with fixed Stueckelberg symmetry is used in~\cite{Bergshoeff:2023fcf}. However, it turns out that it is easier to understand the quantum mechanics of the worldsheet theory using the $(v_+\,,\,u_\text{M})$ parametrization.  

\paragraph{\emph{Yang-Mills gauge anomalies.}}
We now analyze the YM gauge anomalies in the path integral associated with the heterotic non-relativistic string action~\eqref{eq:nrhst123}. This is essentially analogous to the standard case that we have discussed in Section~\ref{sec:aolea}, except that the coupling between the current $J^{AB}_z = i \, \lambda^A \, \lambda^B$ associated with the left-moving fermions and the pullback of the gauge potential $V_{\bar{z}}$ in Eq.~\eqref{eq:inta} is now replaced with
\begin{equation}
	S_\text{int} = \frac{1}{4\pi} \int \dd^2 z \, \tr \Bigl[ J^{}_z \, \bigl( u^{}_{\bar{z}} - \chi^{}_- \, v^{}_+ \bigr) \Bigr]\,,
        \qquad%
    u^{}_{\bar{z}} \equiv \p^{}_{\bar{z}} X^\text{M} \, u^{}_\text{M}\,.
\end{equation}
Varying the path integral and focusing on the lowest-order quantum corrections, we find that the gauge transformation of the path integral leads to the anomalous term
$$
    \delta_\xi \CZ \sim - \frac{1}{8\pi^2} \int \frac{\dd z_1^2 \, \dd z_2^2}{z_{12}^2} \, \tr \biggl\{ \Bigl[ u^{}_{\bar{z}^{}_1} (\bar{z}^{}_1) - \chi^{}_- (\bar{z}^{}_1) \, v^{}_+ (\bar{z}^{}_1) \Bigr] \, \p^{}_{\bar{z}^{}_2} \xi (\bar{z}^{}_2) \biggr\}. 
$$
After integrating by parts and applying the identity $\p_{\bar{z}} z^{-2} = -2 \pi \, \delta^{(2)} (z\,,\,\bar{z})$\,, we find that Eq.~\eqref{eq:dxz} is now replaced with
\begin{equation} \label{eq:ganrs}
	\delta_\xi \mathcal{Z} \sim \frac{1}{4\pi} \int \dd^2 z \, \tr \Bigl[ \partial_z \xi \, \bigl(u^{}_{\bar{z}} - \chi^{}_- \, v^{}_+ \bigr) \Bigr]. 
\end{equation}
In order to cancel this anomalous term, we modify $\CE_\text{MN}$ in Eq.~\eqref{eq:nrhst123} to be
\be \label{eq:ceap}
    \CE^{}_\text{MN} = E^{}_\text{MN} + b^{}_\text{MN} + \frac{\alpha'}{4} \, \tr \bigl( u^{}_\text{M} \, u^{}_\text{N} \bigr)\,,
\ee
and furthermore require that $b^{}_\text{MN}$ and $\tau^{}_\text{M}{}^-$ transform under the YM gauge symmetry as
\begin{align} \label{eq:apymg}
    \delta^{}_\xi b^{}_\text{MN} = - \frac{\alpha'}{4} \, \tr \Bigl[ \xi \, \bigl( \p^{}_{\text{M}} u^{}_{\text{N}} - \p^{}_{\text{N}} u^{}_{\text{M}} \bigr) \Bigr],
        \qquad%
    \delta^{}_\xi \tau^{}_\text{M}{}^- = - \frac{\alpha'}{2} \, \tr \bigl( v^{}_+ \, \p^{}_\text{M} \xi \bigr)\,,
\end{align}
where we have recovered the $\alpha'$ dependence. 
While the gauge transformation of $b^{}_\text{MN}$ is akin to Eq.~\eqref{eq:dxib}, the gauge transformation of $\tau^{}_\text{M}{}^-$ in the above equation is required to cancel the contribution in Eq.~\eqref{eq:ganrs} that is proportional to $\chi^{}_-$\,. 
Combined with the remaining gauge transformations in Eq.~\eqref{eq:gthsnr}, we conclude that the gauge anomaly~\eqref{eq:ganrs} is canceled in the modified path integral at the lowest order in $\alpha'$\,. 
Finally, in order to compensate the boost transformation~\eqref{eq:dgum} of $u^{}_\text{M}$ in the generalized metric~\eqref{eq:ceap}, the string Galilean boost transformation of the transverse vielbein $E_\text{M}{}^i$ in Eq.~\eqref{eq:sgb} now also acquires an $\alpha'$ dependence, with 
\be \label{eq:dgeim}
    \delta^{}_\text{G} E^{}_\text{M}{}^{i} = \Lambda^{}_I{}^{i} \, \tau^{}_\text{M}{}^I + \frac{\alpha'}{4} \, \Lambda^{}_-{}^i \, \tr \bigl( v^{}_+ \, u^{}_\text{M} \bigr)\,.
\ee
Here, the $\alpha'$ dependence in Eq.~\eqref{eq:dgeim} is required to compensate the boost transformation~\eqref{eq:dgum} of $u^{}_\text{M}$ in the generalized metric~\eqref{eq:ceap}. 

\paragraph{\emph{Back to the DLCQ.}}
In the above, we deduced the heterotic non-relativistic string sigma model~\eqref{eq:nrhst123} from a lightlike T-duality transformation of conventional heterotic string theory in the DLCQ. Now that we have formulated the heterotic non-relativistic string, we will in turn use the new sigma model~\eqref{eq:nrhst123} to define heterotic string theory in the DLCQ via a T-duality transformation along a spatial circle. 
Consider a longitudinal spacelike Killing vector $k^\text{M}$ satisfying
$k^\text{M} \, \tau^{}_\text{M}{}^0 = 0$\,,
$k^\text{M} \, \tau^{}_\text{M}{}^1 \neq 0$\,,
and
$k^\text{M} \, E^{}_\text{M}{}^i = 0$\,.
Note that we have used the longitudinal and transverse vielbein fields to project the Killing vector, such that we are able to specify the frame index with which the Killing vector is aligned. 
Next, we introduce the adapted coordinates $X^\text{M} = (X^\mu\,, \, x)$ with $\p_x = k^\text{M} \, \p_\text{M}$\,, such that
\be
	\tau^{}_x{}^0 = 0\,,
		\qquad%
	\tau^{}_x{}^1 \neq 0\,,
		\qquad%
	E^{}_x{}^i = 0\,.
\ee
We now T-dualize the heterotic non-relativistic string sigma model~\eqref{eq:nrhst123}, where the background field $\CE_\text{MN}$ is given by~\eqref{eq:ceap}, in the longitudinal spacelike isometry $x$\,. 
Following~\cite{Bergshoeff:2018yvt}, we find that the dual action is
\be
    \tilde{S} = - \frac{1}{4\pi\alpha'} \int \dd^2 \sigma \, \biggl[ \p^{}_z X^\text{M} \, \p^{}_{\bar{z}} X^\text{N} \, G^{}_{\text{M}\text{N}} + \tr \Bigl( \lambda \, \p^{}_{\bar{z}} \lambda - i \, \lambda \, V^{}_{\text{M}} \, \p^{}_{\bar{z}} X^\text{M} \, \lambda \Bigr) \biggr]\,,
\ee
where
\begin{subequations} \label{eq:hbrslstd}
\begin{align}
    G^{}_{xx} &= 0\,, 
        \qquad%
    G^{}_{x \mu} = \frac{\tau^{}_\mu{}^+}{\tau_x{}^1}\,,
        \qquad%
    G^{}_{\mu x} = \frac{\tau^{}_\mu{}^-}{\tau_x{}^1}\,, 
        \qquad%
    V^{}_x = - \frac{v_+}{2 \, \tau_x{}^1}\,, \\[4pt]
    G^{}_{\mu\nu} &= \CE^{}_{\mu\nu} - \frac{\tau^{}_\mu{}^- \, \tau^{}_\nu{}^+}{\bigl( \tau_x{}^1 \bigr)^2} \, \CE^{}_{xx} - \frac{1}{\tau_x{}^1} \, \Bigl( \tau^{}_\mu{}^- \, \CE^{}_{x\nu} - \tau^{}_\nu{}^+ \, \CE^{}_{\mu x} \Bigr)\,, \\[4pt]
    V^{}_\mu &= u^{}_\mu + \frac{\tau_\mu{}^+}{\tau_x{}^1} \biggl( u^{}_x + \frac{\CE_{xx}}{\tau_x{}^1} \, v^{}_+ \biggr) + \frac{v_+}{\tau_x{}^1} \, \CE^{}_{x\mu}\,,
        \qquad%
    \Phi = \phi - \ln \bigl| \tau_x{}^1 \bigr|\,.
\end{align}
\end{subequations}
In the T-dual frame, the target space acquires a 10D background metric with a Lorentzian structure group. 
This dual action describes the usual heterotic string in the DLCQ, as $G_{xx}$ vanishes identically. 
The Buscher rules~\eqref{eq:hbrslstd} are consistent with the ones found in~\cite{Bergshoeff:2023fcf}. 

In the next section, we will show how the heterotic non-relativistic string arises from a well-defined BPS decoupling limit of the conventional heterotic string, and is therefore believed to be a self-contained theory. Via the spatial T-duality transformation that we have discussed above, heterotic non-relativistic string theory may be used as a first-principles definition of the DLCQ of heterotic string theory. This is a heterotic generalization of the discussion in~\cite{Bergshoeff:2018yvt}, where it is shown that non-relativistic string theory, as a self-contained non-relativistic quantum gravity, provides a first-principles definition for the DLCQ of conventional string theory. The analogs of transverse and lightlike T-duality transformations of non-relativistic string theory considered in~\cite{Bergshoeff:2018yvt} can also be generalized to the heterotic case. We will discuss this in detail in Appendix~\ref{app:tdhnrst}.

\section{From Current-Current Deformation to BPS Decoupling Limit} \label{sec:ccdbps}

In the previous section we have constructed the heterotic non-relativistic string action~\eqref{eq:nrhst123} by T-dualizing relativistic heterotic string theory in the DLCQ. In this section we show how heterotic non-relativistic string theory arises from a BPS decoupling limit of relativistic heterotic string theory. In Section~\ref{sec:ttdnrrst} we will review how to obtain the Gomis-Ooguri action~\eqref{eq:dualnrst}, which underlies type II non-relativistic string theory, from a BPS decoupling limit zooming in on the fundamental string. It is shown in~\cite{Blair:2020ops, Blair:2024aqz} that the inverse of such a limit in the Nambu-Goto formulation is the $T\bar{T}$ deformation, which turns non-relativistic string theory into the relativistic one. We will then discuss in Section~\ref{sec:ccdhst} how similar techniques can be applied to derive the BPS decoupling limit associated with the heterotic non-relativistic string theory, which is in agreement with the supergravity derivation in~\cite{Bergshoeff:2023fcf}. Finally, we will generalize this limiting procedure to construct the $\CN = 1$ supersymmetric sigma models for the heterotic non-relativistic string and analyze the gravitational anomalies. 

\subsection{\texorpdfstring{$T\bar{T}$}{TTbar}-Deformation from Non-Relativistic to Relativistic String Theory} \label{sec:ttdnrrst}

In order to develop a systematic way to construct the limiting procedure that leads to heterotic non-relativistic string theory, we start with the simpler type II case, whose essence is captured by the bosonic Gomis-Ooguri action~\eqref{eq:dualnrst}, which we rewrite as follows
\begin{align} \label{eq:goant}
    S^{}_\text{NRST} = \frac{1}{4\pi} \int \dd^2 z \, \Bigl( \p^{}_{\!z} X^\text{M} \, \p^{}_{\bar{z}} X^\text{N} \, \CE^{}_{\text{M}\text{N}} + \chi^{}_+ \, \p^{}_{\bar{z}} X^\text{M} \, \tau^{}_\text{M}{}^+ + \chi^{}_- \, \p^{}_{z} X^\text{M} \, \tau^{}_\text{M}{}^- \Bigr)\,,
\end{align}
where the tildes in Eq.~\eqref{eq:dualnrst} are dropped. Recall that $\CE_\text{MN} = E_\text{M}{}^i \, E_\text{N}{}^i + b_\text{MN}$\,, $i = 2\,,\, \cdots, \, 9$\,. The target space has a string Galilei symmetry~\cite{Brugues:2004an, Brugues:2006yd, Andringa:2012uz}, which consists of longitudinal and transverse translations, longitudinal Lorentz boost, transverse spatial rotations, and the string Galilei boosts~\eqref{eq:sgb}.\,\footnote{See also~\cite{Bidussi:2021ujm} for the modern perspective of F-string Galilei algebra.} We denote the generators associated with the transverse translations and the string Galilei boosts as $P_i$ and $G_{Ij}$\,, where $I = 0\,,\,1$ is the longitudinal index. In the string Galilei algebra, we have $[P_i\,, \, G_{Ij}] = 0$\,, and the realization of this target space symmetry does not require any further constraints on the background fields. Quantum mechanically, the following marginal deformation will be generated~\cite{Gomis:2019zyu, Gallegos:2019icg, Yan:2019xsf, Yan:2021lbe}:
\be \label{eq:ccdef}
    - \frac{1}{4\pi\alpha'} \int \dd^2\sigma \, \mathbf{t} (X) \, \chi^{}_+ \, \chi^{}_-\,,
\ee
where $\mathbf{t}(X)$ is a new coupling.\,\footnote{From the supergravity perspective, this $\mathbf{t}$ is a Stueckelberg field, which can be absorbed into other background fields.} It is however possible to prevent the operator from being generated quantum mechanically by starting with a stronger symmetry principle, by introducing an extension of the string Galilei algebra in which 
\be \label{eq:pgz}
    [P_i\,, \, G_{jI}] = \delta_{ij} \, Z_I\,,
\ee
where $Z_I$ is the analog of the generator in the Bargmann algebra associated with the particle number conservation. This $Z_I$ generator is associated with the string winding conservation in non-relativistic string theory. A non-trivial realization of the extra $Z_I$ symmetry in the Gomis-Ooguri action requires that $\tau^{}_\text{M}{}^I$ satisfy the torsional constraint~\cite{Andringa:2012uz} 
\be \label{eq:tcf}
    \nabla^{}_{\![\text{M}} \tau^{}_{\text{N}]}{}^I \equiv \p^{}_{[\text{M}} \tau^{}_{\text{N}]}{}^I - \epsilon^I{}_J \, \omega^{}_{[\text{M}} \, \tau^{}_{\text{N}]}{}^J = 0\,,
\ee
where $\omega^{}_\text{M}$ is the spin connection that is dependent of the vielbein fields and is associated with the longitudinal Lorentz boost. 
Note that Eq.~\eqref{eq:tcf} is only an intrinsic geometric constraint after projecting out the spin connection using the inverse transverse vielbein fields. 

It is shown in~\cite{Yan:2021lbe} that the operator~\eqref{eq:ccdef} is not generated perturbatively at any loop order as long as the torsionless condition~\eqref{eq:tcf} on $\tau^{}_\text{M}{}^I$ is imposed. Moreover, it is likely that the same condition (together with accompanying conditions on various gauge field strengths) may be required for the self-consistency of the non-Lorentzian supergravity in type II non-relativistic superstring theory~\cite{Bergshoeff:2024nin}. It is alleged that type II non-relativistic superstring theory is a self-consistent corner with emergent symmetries. Turning on the `torsional' deformation~\eqref{eq:ccdef} makes the theory flow away from the non-relativistic string corner towards type II string theory.

Note that the operator~\eqref{eq:ccdef} is the only extra marginal deformation that can be turned on in the sigma model. Moreover, the $X^\text{M}$-dependence in $\mathbf{t}(X)$ can be absorbed into $\tau_\text{M}{}^\pm$ by rescaling $\chi_\pm$\,, such that $\mathbf{t}$ is constant. This phenomenon is the origin of the emergent dilatational symmetry in non-relativistic string theory~\cite{Bergshoeff:2018yvt, Bergshoeff:2019pij, Bergshoeff:2021tfn, Yan:2021lbe}, which we explain below. For simplicity, we assume that $\mathbf{t} > 0$\,, which in type IIB non-relativistic superstring theory can be viewed as choosing a particular branch in the SL($2,\mathbb{Z}$) S-duality group~\cite{Bergshoeff:2022iss}. Note that the deformed action that combines Eqs.~\eqref{eq:dualnrst} and~\eqref{eq:ccdef}, 
\be \label{eq:defrel}
    S^{}_\text{def.} = \frac{1}{4\pi} \int \dd^2 z \, \Bigl( \p^{}_{\!z} X^\text{M} \, \p^{}_{\bar{z}} X^\text{N} \, \CE^{}_{\text{M}\text{N}} + \chi^{}_+ \, \p^{}_{\bar{z}} X^\text{M} \, \tau^{}_\text{M}{}^+ + \chi^{}_- \, \p^{}_{z} X^\text{M} \, \tau^{}_\text{M}{}^- + \mathbf{t} \, \chi^{}_+ \, \chi^{}_- \Bigr),
\ee
is invariant under the following transformations parametrized by $\Delta = \Delta (X)$\,, 
\be \label{eq:ds}
    \chi^{}_\pm \rightarrow \Delta \, \chi^{}_\pm\,,
        \qquad%
    \tau^{}_\text{M}{}^\pm \rightarrow \frac{\tau^{}_\text{M}{}^\pm}{\Delta}\,,
        \qquad%
    \mathbf{t} \rightarrow \frac{\mathbf{t}}{\Delta^2}\,. 
\ee
Moreover, this dilatational transformation also shifts the dilaton by $\ln \Delta$\,. 
In the $\mathbf{t} \rightarrow 0$ limit, the $\mathbf{t}$-dependent term drops out and non-relativistic string theory arises, which is invariant under the rescaling~\eqref{eq:ds} of the longitudinal vielbein $\tau^{}_\text{M}{}^\pm$. This is the emergent dilatational symmetry in non-relativistic string theory. Now, taking $\Delta^2 = \omega^2 \, \mathbf{t}$ for a constant $\omega$\,, we are essentially led to a representation where $\mathbf{t}$ is fixed to a constant value with
\be
    \mathbf{t} \rightarrow \frac{1}{\omega^2}\,.
\ee
Integrating out $\chi^{}_\pm$ in Eq.~\eqref{eq:defrel}, we are led to the standard Polyakov string action,
\be
    S^{}_\text{rel.} = \frac{1}{4\pi} \int \dd^2 z \, \p^{}_{\!z} X^\text{M} \, \p^{}_{\bar{z}} X^\text{N} \, \bigl( g^{}_\text{MN} + B^{}_\text{MN} \bigr)\,,
\ee
where the background metric $g^{}_\text{MN}$ and Kalb-Ramond field $B^{}_\text{MN}$ are reparametrized as~\cite{Bergshoeff:2019pij}
\begin{align} \label{eq:nrstpre}
    g^{}_\text{MN} = \omega^2 \, \tau^{}_\text{M}{}^I \, \tau^{}_\text{N}{}^J \, \eta^{}_{IJ} + E^{}_\text{MN}\,, 
        \qquad%
    B^{}_\text{MN} = - \omega^2 \, \tau^{}_\text{M}{}^I \, \tau^{}_\text{N}{}^J \, \epsilon^{}_{IJ} + b_\text{MN}\,,
\end{align}
Here, $I = 0\,, \, 1$ and $\tau^{}_\text{M}{}^\pm = \tau^{}_\text{M}{}^0 \pm \tau^{}_\text{M}{}^1$\,. Moreover, the dilaton $\Phi$ in relativistic string theory also acquires a reparametrization as $\Phi = \varphi + \ln \omega$~\cite{Bergshoeff:2019pij}, with $\varphi$ the effective dilaton for the resulting non-relativistic string. The $\omega \rightarrow \infty$ limit of (type II) string theory parametrized  as in Eq.~\eqref{eq:nrstpre} gives rise to non-relativistic string theory. This is a BPS decoupling limit: we zoom in on a background fundamental string configuration by fine tuning the string charge associated with the $B$-field to cancel the infinite string tension. In the resulting non-relativistic string theory, the light excitations are the winding closed non-relativistic strings, and they interact with each other via an instantaneous Newton-like gravitational force~\cite{Gomis:2000bd, Danielsson:2000mu}.  

The deformation~\eqref{eq:ccdef} can be thought of as a current-current deformation of the non-relativistic string sigma model~\eqref{eq:goant}.  
Define $P^{}_\text{M} = \p \CL / \p X^\text{M}$ to be the momentum conjugate to the worldsheet fields $X^\text{M}$, where $\CL$ is the Lagrangian associated with the Gomis-Ooguri action~\eqref{eq:goant}. It then follows that $\chi^{}_\pm = \tau^\text{M}{}_\pm \, P^{}_\text{M}$ are projections of the conjugate momentum onto the longitudinal light-cone directions. The deformation~\eqref{eq:ccdef} is therefore a marginal operator that is bi-linear in the momentum currents.  

In the Nambu-Goto formulation, such a current-current deformation is mapped to the standard $T\bar{T}$-deformation~\cite{Blair:2020ops, Blair:2024aqz}. Using Eq.~\eqref{eq:cov} to undo the conformal gauge in the deformed action~\eqref{eq:defrel}, we find
$$
    S^{}_\text{def.} = \frac{1}{8\pi} \int \dd^2 
    \sigma\, e \, \Bigl( \CD X^\text{M} \, \bar{\CD} X^\text{N} \, \CE^{}_{\text{M}\text{N}} + \chi^{}_+ \, \bar{\CD} X^\text{M} \, \tau^{}_\text{M}{}^+ + \chi^{}_- \, \CD X^\text{M} \, \tau^{}_\text{M}{}^- + \mathbf{t} \, \chi^{}_+ \, \chi^{}_- \Bigr)\,.
$$
Here, we used the notation introduced at the end of Section~\ref{sec:tdt}. Integrating out the worldsheet zweibein $e^{}_\alpha{}^a$ gives rise to the Nambu-Goto action. In flat background fields with $\tau^{}_\text{M}{}^I = \delta_\text{M}^I$\,, $E^{}_\text{M}{}^i = \delta_\text{M}^i$\,, and $b^{}_\text{MN} = 0$, we Wick rotate the Euclidean time on the worldsheet to the real time and then take the static gauge with $X^{0\,,\,1} = \sigma^{0\,,\,1}$\,. In this way we obtain the following  Nambu-Goto type Lagrangian:
\be \label{eq:nga}
    \CL (\mathbf{t}) = - \frac{1}{4\pi} \, \frac{1}{\mathbf{t}} \, \biggl[ \sqrt{- \det \Bigl( \eta^{}_{\alpha\beta} + \mathbf{t} \, \p^{}_\alpha X^i \, \p^{}_\beta X^i \Bigr)} - 1 \biggr]\,.
\ee
In the $\mathbf{t} \rightarrow 0$ limit, we are led to the `seed' Lagrangian, 
\be \label{eq:seedl}
    \CL (0) = - \frac{1}{8\pi} \, \p_\alpha X^i \, \p^\alpha X^i\,,
\ee
which describe 8 free bosons. 
The deformed Lagrangian $\CL(\mathbf{t})$ arises from the $T\bar{T}$ deformation of $\CL(0)$ and satisfies the flow equation~\cite{Zamolodchikov:2004ce,Smirnov:2016lqw,Cavaglia:2016oda,Bonelli:2018kik},
\be \label{eq:floweqn}
    \frac{\dd\CL(\mathbf{t})}{\dd\mathbf{t}} \sim \det \bigl[ T_{\alpha\beta} (\mathbf{t}) \bigr]\,, 
\ee
where $T_{\alpha\beta} (\mathbf{t})$ is the stress-energy tensor associated with the deformed Lagrangian $\CL (\mathbf{t})$\,. The marginal operator~\eqref{eq:ccdef} in the Polyakov formulation is now mapped to an infinite series of irrelevant operators that $T\bar{T}$-deform the seed theory~\eqref{eq:seedl} to Eq.~\eqref{eq:nga}. 

In curved background fields, the deformed Nambu-Goto Lagrangian~\eqref{eq:nga} now takes the standard form as in relativistic string theory (in real time on the worldsheet), 
\be
    \CL(\mathbf{t}) = - \frac{1}{4\pi} \biggl[ \sqrt{- \det \Bigl( \p^{}_\alpha X^\text{M} \, \p^{}_\beta X^\text{N} \, g^{}_\text{MN}\Bigr)} + \tfrac{1}{2} \, \epsilon^{\alpha\beta} \, \p_\alpha X^\text{M} \, \p_\beta X^\text{N} \, B^{}_\text{MN} \biggr]\,, 
\ee
where the metric $g^{}_\text{MN}$ and the Kalb-Ramond field $B^{}_\text{MN}$ are reparametrized as in Eq.~\eqref{eq:nrstpre}.
In the $\mathbf{t} \rightarrow 0$ (\emph{i.e.}, $\omega \rightarrow \infty$) limit, the seed Lagrangian~\eqref{eq:seedl} becomes
\be
    \CL(0) = - \frac{1}{8\pi} \, \Bigl( \sqrt{-\det \tau} \, \tau^{\alpha\beta} \, E_\text{MN} + \epsilon^{\alpha\beta} \, b^{}_\text{MN} \Bigr) \, \p_\alpha X^\text{M} \, \p_\beta X^\text{N}\,,
\ee
with $\tau^{}_{\alpha\beta} = \p^{}_\alpha X^\text{M} \, \p^{}_\beta X^\text{N} \, \tau^{}_\text{MN}$ the pullback of $\tau^{}_\text{MN} \equiv \tau^{}_\text{M}{}^I \, \tau^{}_\text{N}{}^J \, \eta^{}_{IJ}$ and $\tau^{\alpha\beta}$ its inverse. After taking the static gauge such that $\tau_{\alpha\beta}$ becomes effectively a worldsheet metric, it is then natural to define the stress energy tensor to be
\be
    T_{\alpha\beta} (\mathbf{t}) = - \frac{2}{\sqrt{-\det \tau}} \frac{\p \CL(\mathbf{t})}{\p \tau^{\alpha\beta}}\,.
\ee
Now, the flow equation~\eqref{eq:floweqn} is generalized to be
\be
    \frac{\p\CL}{\p\mathbf{t}} \sim \sqrt{-\det \tau} \, \det \bigl[ T_{\alpha\beta} (\mathbf{t}) \bigr]\,. 
\ee
Therefore, in the context of string theory, the $T\bar{T}$ deformation can be viewed as the inverse of the BPS decoupling limit that zooms in on a background fundamental string in type II superstring theory, and it is equivalent to the marginal current-current deformation~\eqref{eq:ccdef} that deforms non-relativistic to relativistic string theory~\cite{Blair:2020ops, Yan:2021lbe, Blair:2024aqz}. 

In the heterotic case, it is difficult to apply the concept of the $T\bar{T}$-deformation as in the type II case, as there does not exist a simple heterotic Nambu-Goto formulation. Nevertheless, it is still possible to consider the current-current deformation~\eqref{eq:ccdef} of the heterotic non-relativistic string, which we study below. 

\subsection{Current-Current Deformation in Heterotic String Theory} \label{sec:ccdhst}

We now apply the current-current deformation~\eqref{eq:ccdef} to heterotic non-relativistic string theory, whose worldsheet action is given in Section~\ref{sec:rgosm}. Combining Eqs.~\eqref{eq:ccdef} and \eqref{eq:nrhst123}, we are led to the following deformed action:
\begin{align} \label{eq:nrhst1def}
\begin{split}
	S = \frac{1}{4\pi} \int & \dd^2 \sigma \, \biggl\{ \p^{}_z X^\text{M} \, \p^{}_{\bar{z}} X^\text{N} \, \CE^{}_{\text{M}\text{N}} + \tr \Bigl( \lambda \, \p^{}_{\bar{z}} \lambda - i \, \lambda \, u^{}_{\text{M}} \, \p^{}_{\bar{z}} X^\text{M} \, \lambda \Bigr) \\[4pt]
    & + \chi^{}_+ \, \p^{}_{\bar{z}} X^{\text{M}} \, \tau^{}_{\text{M}}{}^+ + {\chi}^{}_- \, \Bigl[ \p^{}_z X^\text{M} \, {\tau}^{}_{\text{M}}{}^- + i \, \tr \bigl( \lambda \, v^{}_+ \, \lambda \bigr) \Bigr] + \omega^{-2} \, \chi^{}_+ \, \chi^{}_- \biggr\}\,,
\end{split}
\end{align}
where we have set $\mathbf{t} = \omega^{-2}$ and ignored the gravitational anomaly. Moreover,
\be \label{eq:ceaps}
    \CE_\text{MN} = E_\text{M}{}^i \, E_\text{N}{}^i + b_\text{MN} + \frac{1}{2} \, \tr \bigl( u^{}_\text{M} \, u^{}_\text{N} \bigr)\,.
\ee
Integrating out the one-form fields $\chi_\pm$ leads to the conventional heterotic string sigma model,
\be
    S = \frac{1}{4\pi} \int \dd^2 \sigma \, \biggl[ \p^{}_z X^\text{M} \, \p^{}_{\bar{z}} X^\text{N} \, G^{}_{\text{M}\text{N}} + \tr \Bigl( \lambda \, \p^{}_{\bar{z}} \lambda - i \, \lambda \, V^{}_{\text{M}} \, \p^{}_{\bar{z}} X^\text{M} \, \lambda \Bigr) \biggr]\,,
\ee
where
\begin{subequations} \label{eq:repa0}
\be 
	{G}^{\phantom{\dagger}}_\text{MN} = - \omega^2 \, \tau^{\phantom{\dagger}}_\text{M}{}^- \, \tau^{\phantom{\dagger}}_\text{N}{}^+ + \CE^{\phantom{\dagger}}_\text{MN}\,,
		\qquad%
	{V}^{\phantom{\dagger}}_\text{M} = \omega^2 \, v^{\phantom{\dagger}}_+ \, \tau^{\phantom{\dagger}}_\text{M}{}^+ + u^{\phantom{\dagger}}_\text{M}\,.
\ee
In addition, the dilaton background should be reparametrized as in~\cite{Bergshoeff:2019pij}, with
\be
    \Phi = \varphi + \ln \omega\,.  
\ee
\end{subequations}
These prescriptions determine how heterotic non-relativistic string theory arises from a BPS decoupling limit of the conventional heterotic string theory. 

\paragraph{\emph{Reparametrizations in components.}} Recall that the generalized metric $G_\text{MN}$ in heterotic string sigma model is
\begin{equation} \label{eq:mms}
    G^{}_\text{MN} = g^{}_\text{MN} + B^{}_\text{MN} + \frac{1}{2} \, \tr \bigl( V^{}_\text{M} \, V^{}_\text{N} \bigr)\,.
\end{equation}
Combining Eqs.~\eqref{eq:repa0} and \eqref{eq:mms}, and in terms of the vielbein fields defined via 
\be
    g^{}_\text{MN} = - E^{}_{(\text{M}}{}^+ \, E^{}_{\text{N})}{}^- + E^{}_\text{M}{}^i \, E^{}_\text{N}{}^i\,,
\ee
we are led to the following limiting prescription:
\begin{subequations} \label{eq:eebrep}
\begin{align}
    E^{}_\text{M}{}^+ &= \omega \, \tau^{}_\text{M}{}^+\,, 
        &%
    B^{}_\text{MN} &= - \omega^2 \, \tau^{}_{[\text{M}}{}^- \, \tau^{}_{\text{N}]}{}^+ + b^{}_\text{MN}\,, \\[4pt]
    E^{}_\text{M}{}^- &= \tfrac{1}{2} \, \omega^3 \, \tr \bigl( v_+^2 \bigr) \, \tau^{}_\text{M}{}^+ + \omega \, \Bigl[ \tau^{}_\text{M}{}^- + \tr( v^{}_+ \, u^{}_\text{M} \bigr) \Bigr]\,,
        &%
    {V}^{\phantom{\dagger}}_\text{M} &= \omega^2 \, v^{\phantom{\dagger}}_+ \, \tau^{\phantom{\dagger}}_\text{M}{}^+ + u^{\phantom{\dagger}}_\text{M}.
\end{align}
\end{subequations}
In the type II case where the gauge bosons $v_+$ and $u_\text{M}$ are both zero (which secretly carries a $\sqrt{\alpha'/2}$ prefactor), the prescription~\eqref{eq:eebrep} reduces to the limiting description in the IIA case given in Eq.~\eqref{eq:nrstpre}.  

The prescription~\eqref{eq:eebrep} is seemingly distinct from the one constructed in~\cite{Bergshoeff:2023fcf}, where the latter was achieved by requiring self-consistency of the decoupling limit in supergravity and T-duality transformations. Intriguingly, these two different limiting prescriptions are related to each other via an involved redefinition of the longitudinal vielbein field. In terms of the new longitudinal vielbein fields denoted by variables with tildes, we write
\begin{align}
    \tau^{}_\text{M}{}^+ &= \tilde{\tau}^{}_\text{M}{}^-\,,
        \qquad%
    E^{}_\text{M}{}^i = \tilde{E}^{}_\text{M}{}^i\,,
        \qquad%
    u^{}_\text{M} = \tilde{u}^{}_- \, \tilde{\tau}^{}_\text{M}{}^- + \tilde{v}^{}_+ \, \tilde{\tau}^{}_\text{M}{}^+ + \tilde{v}^{}_i \, \tilde{E}^{}_\text{M}{}^i\,,
        \qquad%
    v^{}_+ = \tilde{v}^{}_- - \frac{\tilde{u}^{}_-}{\omega^2}\,, \notag \\[2pt]
    \tau^{}_\text{M}{}^- &= \Bigl[ 1 - \tr\bigl(\tilde{v}^{}_+ \, \tilde{v}^{}_- \bigr) \Bigr] \, \tilde{\tau}^{}_\text{M}{}^+ - \tr\bigl( \tilde{v}^{}_- \, \tilde{v}^{}_i \bigr) \, \tilde{E}^{}_\text{M}{}^i + \frac{\tilde{u}^{}_- \, \tilde{\tau}^{}_\text{M}{}^- + 2 \, \tilde{v}^{}_+ \, \tilde{\tau}^{}_\text{M}{}^+ + 2 \, \tilde{v}^{}_i \, \tilde{E}^{}_\text{M}{}^i}{2 \, \omega^2}\,, \\[8pt]
    b^{}_\text{MN} &= \tfrac{1}{2} \, \tilde{b}^{}_\text{MN} - \tr \bigl( \tilde{u}^{}_- \, \tilde{v}^{}_i \bigr) \, \tilde{\tau}^{}_{[\text{M}}{}^- \, \tilde{E}^{}_{\text{N}]}{}^i - \tr \bigl( \tilde{u}^{}_+ \, \tilde{v}^{}_+ \bigr) \, \tilde{\tau}^{}_{[\text{M}}{}^+ \, \tilde{\tau}^{}_{\text{N}]}{}^-\,. \notag 
\end{align}
Consequently, the prescriptions in Eqs.~\eqref{eq:ceaps} and \eqref{eq:eebrep} become
\begin{subequations}
\begin{align}
    \tilde{E}^{}_\text{M}{}^- &\equiv E^{}_\text{M}{}^+ = \omega \, \tilde{\tau}^{}_\text{M}{}^-\,, 
        \qquad%
    \tilde{E}^{}_\text{M}{}^+ \equiv E^{}_\text{M}{}^- = - \frac{1}{2} \, \omega^3 \, \tilde{v}^{AB}_- \, \tilde{v}^{AB}_- \, \tilde{\tau}^{}_\text{M}{}^- + \omega \, \tilde{\tau}^{}_\text{M}{}^+\,, \\[4pt]
    \tilde{B}^{}_\text{MN} &\equiv 2 \, B^{}_\text{MN} = \omega^2 \, \Big( 1 + \tilde{v}^{AB}_+ \, \tilde{v}^{AB}_- \Big) \, \tilde{\tau}^{}_\text{M}{}^I \, \tilde{\tau}^{}_\text{N}{}^J \, \epsilon^{}_{IJ} + 2 \, \omega^2 \, \tilde{v}^{AB}_- \, \tilde{v}^{AB}_i \, \tilde{\tau}^{}_{[\text{M}}{}^- \, \tilde{E}^{}_{\text{N}]}{}^i + \tilde{b}^{}_\text{MN}\,, \\[4pt]
    \tilde{V}^{}_\text{M} &\equiv V^{}_\text{M} = \omega^2 \, \tilde{v}^{}_- \, \tilde{\tau}^{}_\text{M}{}^- + \tilde{v}^{}_+ \, \tilde{\tau}^{}_\text{M}{}^+ + \tilde{v}^{}_i \, \tilde{E}^{}_\text{M}{}^i\,.
\end{align}
\end{subequations}
which, upon identifying $\omega$ with $c$\,, precisely match Eq.~(7) in~\cite{Bergshoeff:2023fcf}.\,\footnote{See also~\cite{Lescano:2024url, Lescano:2025ixp} for discussions from the double field theoretic (DFT) perspective, where a different limit is proposed, guided by the requirement that the DFT generalized metric remains finite.} Note that $\tilde{\tau}^{}_\text{M}{}^0 = (\tilde{\tau}^{}_\text{M}{}^+ + \tilde{\tau}^{}_\text{M}{}^-)/\sqrt{2}$ and $\tilde{\tau}^{}_\text{M}{}^1 = (\tilde{\tau}^{}_\text{M}{}^+ - \tilde{\tau}^{}_\text{M}{}^-)/\sqrt{2}$\,, which differs from our convention~\eqref{eq:lv}. 

\paragraph{\emph{Gauge symmetry revisited.}} We now derive the YM gauge symmetry in heterotic non-relativistic string theory from the limiting perspective. Before the limit is performed, we consider the YM gauge transformation, which acts on the gauge potentials $B_\text{MN}$ and $V_\text{M}$ in heterotic string theory as in Eqs.~\eqref{eq:gtrnsf} and~\eqref{eq:dxib}. It follows that 
\be
    \delta_\xi G^{}_\text{MN} =  \p^{}_\text{M}\xi \, V^{}_\text{N}\,,
        \qquad%
    \delta_\xi V^{}_\text{M} = \p^{}_{\text{M}} \xi + i \, \bigl[\xi\,,V^{}_\text{M}\bigr]\,,
\ee
Note that the YM gauge transformations are defined up to the  gauge transformations of the $B$-field, \emph{i.e.}~$\delta_\zeta B^{}_\text{MN} = 2 \, \p^{}_{[\text{M}} \zeta^{}_{\text{N}]}$\,.
Projecting the reparametrization of the background fields in Eq.~\eqref{eq:repa0},  using the inverse vielbein fields and taking the limit $\omega \to \infty$\,, we read off that
\begin{subequations}
\begin{align}
    \delta_\xi \tau^{}_\text{M}{}^- &= - v^{}_+ \, \p^{}_\text{M} \xi\,,
        &%
    \delta_\xi v^{}_+ &= i \, [\xi\,,\,v_+]\,, \\[4pt]
    \delta_\xi b^{}_\text{MN} &= - \frac{1}{2} \, \tr \Bigl[ \xi \, \bigl( \p^{}_\text{M} u^{}_\text{N} - \p^{}_\text{N} u^{}_\text{M} \bigr) \Bigr]\,,
        &%
    \delta_\xi u^{}_\text{M} &= \nabla^{}_{\!\text{M}} \xi\,,
\end{align}
\end{subequations}
thereby reproducing the  YM gauge transformations~\eqref{eq:gthsnr} and \eqref{eq:apymg}.

\subsection{Superstring Action and Gravitational Anomalies} \label{sec:satc}

\paragraph{\emph{Superspace formulation.}} Finally, based on the limiting prescription~\eqref{eq:repa0}, the full $\CN=1$ supersymmetric sigma model describing heterotic non-relativistic string theory can be readily derived. We start with the superspace action~\eqref{eq:ssa} describing the conventional heterotic string. 
Now, the same reparametrization~\eqref{eq:repa0} still applies, except that the argument $X^\text{M}$ is now replaced with the superfield $Y^\text{M}$. 
In the $\omega \rightarrow \infty$ limit of the heterotic string action~\eqref{eq:ssa}, we find the following superspace formulation of heterotic non-relativistic string sigma models,
\begin{align} \label{eq:SS2}
\begin{split}
	S = \frac{1}{4\pi} \int \dd^2 z \, \dd\bar{\theta} \, \biggl\{ & \p^{}_z Y^\text{M} \, D^{}_{\bar{\theta}} Y^\text{N} \, \CE^{}_\text{MN} (Y) - \tr \bigl( \Lambda \, \nabla_{\!\bar{\theta}} \Lambda \bigr) \\[4pt]
	& \qquad + \Gamma^{}_+ \, D^{}_{\bar{\theta}} Y^\text{M} \, \tau^{}_\text{M}{}^+ + \Gamma^{}_- \, \Bigl[ \p^{}_z Y^\text{M} \, \tau^{}_\text{M}{}^- + i \, \tr\bigl( \Lambda \, v^{}_+ \, \Lambda \bigr) \Bigr] \biggr\}\,,
\end{split}
\end{align}
where
\be
    \nabla_{\!\bar{\theta}} \Lambda = D_{\bar{\theta}} \Lambda - i \, D_{\bar{\theta}} Y^\text{M} \, u_\text{M} (Y) \, \Lambda\,,
        \qquad%
    \Gamma^{}_+ = \chi^{}_+ - i \, \bar{\theta} \, \gamma^{}_+\,,
        \qquad%
    \Gamma^{}_- = i \, \gamma^{}_- + \bar{\theta} \, \chi^{}_-\,.
\ee
Here, $\chi^{}_\pm$ are bosonic while $\gamma^{}_\pm$ are fermionic. 
Integrating out the supercoordinate $\bar{\theta}$\,, we find 
\begin{align} \label{eq:fa}
\begin{split}
    S = \frac{1}{4\pi} \int & \dd^2 z \, \biggl\{ \p_z X^\text{M} \, \p_{\bar{z}} X^\text{N} \, \CE^{}_\text{MN} + \psi^i \, \nabla_{\!z} \psi^i + \tr \Bigl( \lambda \, \p_{\bar{z}} \lambda - i \, \lambda \, \bar{\p} X^\text{M} \, u^{}_\text{M} \, \lambda \Bigr)  \\[4pt]
    & + \chi_+ \, \Bigl( \bar{\p} X^\text{M} \, \tau^{}_\text{M}{}^+ - \psi^\text{M} \, \psi^\text{N} \, \nabla^{}_{\!\text{M}} \tau^{}_{\text{N}}{}^+ \Bigr) + \chi_- \, \Bigl[ \p X^\text{M} \, \tau^{}_\text{M}{}^- + i \, \tr \bigl( \lambda \, v^{}_+ \, \lambda \bigr) \Bigr] \\[4pt]
    & + \gamma^{}_+ \, \psi^+ 
    + \gamma^{}_- \, \Bigl[ \nabla_{\!z} \psi^- + 2 \, \psi^\text{M} \, \p_z X^{\text{N}} \, \nabla^{}_{[\text{M}} \tau^{}_{\text{N}]}{}^- - i \, \tr \bigl( \lambda \, \psi^\text{M} \, D^{}_{\text{M}} v^{}_+ \, \lambda \bigr) \Bigr] \\[4pt]
    & + \tfrac{i}{2} \, \tr \Bigl( \lambda \, \psi^\text{M} \, \psi^\text{N} \, F^{}_\text{MN} \, \lambda \Bigr) \biggr\}\,, 
\end{split}
\end{align}
where $\psi^\pm \equiv \psi^\text{M} \, \tau^{}_\text{M}{}^\pm$ and $\psi^i \equiv \psi^\text{M} \, \tau^{}_\text{M}{}^i$\,. In order to obtain the above action,  we integrated out the auxiliary field $F^A$ in $\Lambda^A$ and redefined the Lagrange multipliers $\chi_\pm$ and $\gamma_\pm$. Moreover,
\begin{align}
    D^{}_{\text{M}} v^{}_+ = \nabla^{}_{\!\text{M}} \, v^{}_+ + i \, \bigl[ u^{}_\text{M}\,,\, v^{}_+ \bigr]\,,
        \qquad%
    F_\text{MN} = \p_\text{M} u^{}_\text{N} - \p_\text{N} u^{}_\text{M} - i \, [u^{}_\text{M}\,,\, u^{}_\text{N}]\,.
\end{align}
together with $\nabla^{}_{\!\text{M}} \, v^{}_+ = \p^{}_{\text{M}} v^{}_+ - \omega^{}_\text{M} \, v^{}_+$ and
\begin{subequations}
\begin{align}
    \nabla_{\!z} \psi^i &= \p_z \psi^i + \omega_z{}^{i+} \, \psi^- + \Omega_z{}^{ij} \, \psi^j\,, 
        &%
    \Omega^{}_\text{M}{}^{ij} &\equiv \omega^{}_\text{M}{}^{ij} + \tfrac{1}{2} \, h^{}_\text{M}{}^{ij}\,, \\[4pt]
    \nabla_{\!z} \psi^- &= \p_z \psi^- - \omega^{}_z \, \psi^-\,,
        &%
    \nabla^{}_{\![\text{M}} \tau^{}_{\text{N}]}{}^\pm &= \p^{}_{\![\text{M}} \tau^{}_{\text{N}]}{}^\pm \pm \omega^{}_{[\text{M}} \tau^{}_{\text{N}]}{}^\pm\,.
\end{align}
\end{subequations}
Here, $\omega^{}_\text{M}$,\, $\omega^{}_\text{M}{}^{i\pm}$, and $\omega^{}_\text{M}{}^{ij}$ are the spin connections associated with the longitudinal Lorentz boost, string Galilei boost, and transverse rotation symmetry, respectively, and 
$h^{(3)} = \dd b^{(2)}$
is the three-form field strength associated with the Kalb-Ramond field. Note that dropping all terms involving $\psi$ in Eq.~\eqref{eq:fa} recovers the action~\eqref{eq:nrhst123}. 

\paragraph{\emph{Gravitational anomalies.}} We have already considered the YM gauge anomalies in Section~\ref{sec:hnrssm}. This analysis continues to hold for the supersymmetric sigma model~\eqref{eq:fa}. With the superspace action~\eqref{eq:SS2}, we are now able to compute the gravitational anomalies in heterotic non-relativistic string theory. 
We start by collecting the coupling to the pullbacks of the spin connections. In particular, since only the transverse right-moving fermions $\psi^i$ have non-trivial OPEs, with 
$\psi^i (\bar{z}) \, \psi^j (0) \sim \delta^{ij} / \bar{z}$\,,
we only need to focus on the coupling to the current $J_{\bar{z}}^{ij} \equiv \psi^i \, \psi^j$\,. The relevant terms are
\be
	S_\text{int} = \frac{1}{4\pi} \int \dd^2 z \, J_{\bar{z}}^{ij} \, \Bigl( \Omega^{}_{zij} - \chi^{}_+ \, E^{\text{M}}{}^{}_i \, E^{\text{N}}{}^{}_j \, \nabla^{}_{[\text{M}} \tau^{}_{\text{N}]}{}^+ \Bigr)\,.
\ee
Therefore, under the transverse rotation parametrized by $\ell^{ij}$\,, the path integral associated with Eq.~\eqref{eq:fa} contains a gravitational anomaly, 
\be
    \delta_\ell \CZ = - \frac{\alpha'}{4\pi} \int \dd^2 z \, \p^{}_{\bar{z}} \ell^{ij} \Bigl( \Omega^{}_{zij} - \chi^{}_+ \, E^{\text{M}}{}^{}_i \, E^{\text{N}}{}^{}_j \, \nabla^{}_{\![\text{M}} \tau^{}_{\text{N}]}{}^+ \Bigr)\,. 
\ee
We cancel the gravitational anomaly by introducing the following modified version of the two-tensor $\CE_\text{MN}$ in the supersymmetric action~\eqref{eq:fa}, with
\begin{equation} \label{eq:mmvvoonr}
    \CE^{}_\text{MN} = E^{}_\text{M}{}^i \, E^{}_\text{N}{}^i + b^{}_\text{MN} + \frac{\alpha'}{4} \, \biggl[ \tr \Bigl( u^{}_\text{M} \, u^{}_\text{N} \Bigr) + \Omega^{}_\text{M}{}^{ij} \, \Omega^{}_{\text{N}}{}^{ij} \biggr], 
\end{equation}
together with the extra transverse rotations acting on $b_\text{MN}$ and $\tau_\text{M}{}^+$\,, 
\begin{align}
    \delta^{}_\ell b^{}_\text{MN} = - \frac{\alpha'}{4} \, \ell^{}_{ij} \, \Bigl( \p^{}_{\text{M}} \Omega^{}_\text{N}{}^{ij} - \p^{}_{\text{N}} \Omega^{}_\text{M}{}^{ij} \Bigr)\,,
        \qquad%
    \delta^{}_\ell \tau^{}_\text{M}{}^+ = \alpha' \, E^{\text{N}}{}^{}_i \, E^{\text{L}}{}^{}_j \, \nabla^{}_{\![\text{N}} \tau^{}_{\text{L}]}{}^+ \, \p^{}_\text{M} \ell^{ij}\,.
\end{align}

\paragraph{\emph{Torsional constraints.}}
In the heterotic case, it is expected that the type II torsional constraint~\eqref{eq:tcf} reduces to
$\nabla^{}_{\![\text{M}} \tau^{}_{\text{N}]}{}^+ = 0$\,.
Such a chiral torsional constraint also arises from a symmetry principle in the Gomis-Ooguri action~\cite{Yan:2021lbe}: instead of taking $[P_i\,,G_{jI}] = \delta_{ij} \, Z_I$ as in Eq.~\eqref{eq:pgz}, we now extend the string Galilei algebra to incorporate
$[P_i\,,G_{j-}] = \delta_{ij} \, Z_-$ and $[P_i\,,G_{j+}] = 0$\,,
which also leads to a closed algebra. Realizing the $Z_-$ symmetry in the Gomis-Ooguri then leads to the same chiral constraint. It is then proven in~\cite{Yan:2021lbe} that the torsional deformation~\eqref{eq:ccdef} is not generated by quantum corrections at all loops as long as the chiral constraint is imposed. Moreover, it is shown in~\cite{Bergshoeff:2021tfn} that the self-consistency of the supersymmetry transformation in the $\CN=1$ supergravity associated with heterotic non-relativistic string theory requires the same chiral constraint. 

\section{Relation to Heterotic Matrix String Theory} \label{sec:hmst}
 
In this final section, we explain how our study of heterotic non-relativistic string theory is related to matrix theory. We first review how the BFSS matrix theory is related to type IIB matrix string theory and the second quantization of type IIB non-relativistic superstring theory. We will then review studies of heterotic matrix string theory and relate them to the second-quantization of heterotic non-relativistic string theory. See Figure~\ref{fig:rm1} for a road map in the type II case and Figure~\ref{fig:rm2} in the heterotic/type I case. We will also briefly comment on several holographic aspects of heterotic matrix string theory.   

\subsection{Matrix Theory from Type II Superstrings} \label{eq:mtttss}

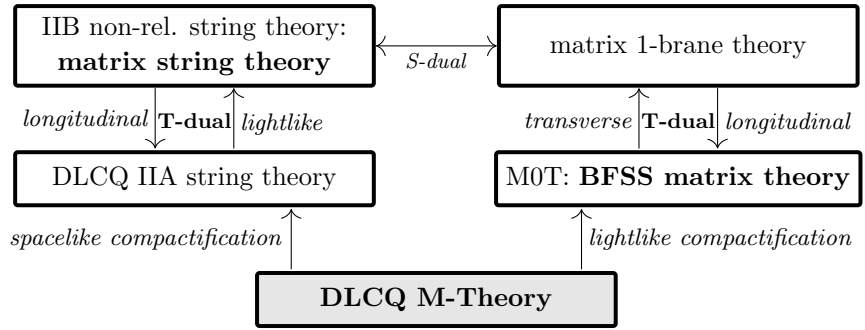
\begin{figure}[b!]
\centering
\begin{tikzpicture}
    \begin{scope}[scale=.8]

        \path[every node/.style=draw, rounded corners=1, line width=1.5pt, minimum width=135pt, minimum height=20pt, font = \small]   
                (3,1.5) node {DLCQ IIA string theory}
                (7,-.5) node[fill=gray!20] {\textbf{DLCQ M-Theory}}
                (11,1.5) node {M0T: \textbf{BFSS matrix theory}}
                (3,3.7) [align=center] node[minimum height = 30pt] {IIB non-rel. string theory:\\
                \textbf{matrix string theory}};

            \path[every node/.style=draw, rounded corners=1, line width=1.5pt, minimum width=135pt, minimum height=30pt, font = \small]
                (11,3.7) [align=center] node {matrix 1-brane theory};

        \path   
                (7,3.25+0.2) [font=\footnotesize] node {\scalebox{0.9}{\emph{S-dual}}}
                (11,2.47) [font=\footnotesize] node {\scalebox{0.9}{\textbf{T-dual}}}
                (1.2,2.47) [font=\footnotesize] node {\emph{longitudinal}}
                (4.45,2.41) [font=\footnotesize] node {\emph{lightlike}}
                (9.35,2.47) [font=\footnotesize] node {\emph{transverse}}
                (12.82,2.415) [font=\footnotesize] node {\emph{longitudinal}}
                (3,2.47) [font=\footnotesize] node {\scalebox{0.9}{\textbf{T-dual}}}
                (3.5-1.3,0.5) [font=\footnotesize] node {\emph{spacelike compactification}}
                (10.3+1.4,0.5) [font=\footnotesize] node {\emph{lightlike compactification}}
                ;

        \draw[{<[length=1.3mm]}-{>[length=1.3mm]}] (6,3.5+0.2) -- (8,3.5+0.2);
        \draw[{[length=1.3mm]}-{>[length=1.3mm]}] (8.5+0.9,0) -- (8.5+0.9,1);
        \draw[{[length=1.3mm]}-{>[length=1.3mm]}] (5.5-0.9,0) -- (5.5-0.9,1);


        \draw [{<[length=1.3mm]}-{[length=1.3mm]}] (2.35,2) -- (2.35,3);
        \draw [{[length=1.3mm]}-{>[length=1.3mm]}] (3.65,2) -- (3.65,3);
        \draw [{[length=1.3mm]}-{>[length=1.3mm]}] (10.35,2) -- (10.35,3);
        \draw [{<[length=1.3mm]}-{[length=1.3mm]}] (11.65,2) -- (11.65,3);
       
\end{scope}
\end{tikzpicture}
\caption{Relations among type II matrix theories and non-relativistic string theory.}
\label{fig:rm1}
\end{figure}

In the original proposal of matrix theory, one considers M-theory compactified over a spatial circle with radius $R_\text{s}$\,. Even though the 11D Lorentz symmetry is already broken due to the spatial compactification, intuitively, it is still useful to think of performing a boost transformation in the spatial circle. In particular, we are ultimately interested in taking the decompactification limit where the 11D Lorentz symmetry is restored. Denote the Lorentz factor by $\gamma$\,. In the infinite boost limit, we send $\gamma$ to infinity and $R_\text{s}$ to zero, while keeping the combination $R \equiv 2 \, \gamma \, R_\text{s}$ fixed. This resulting $R$ then acts as the effective radius of a lightlike compactification. We are therefore led to DLCQ M-theory~\cite{Susskind:1997cw, Seiberg:1997ad, Sen:1997we}, \emph{i.e.}~M-theory on a lightlike compactification, whose dynamics is described by the BFSS matrix theory at finite $N$, with $N$ the Kaluza-Klein momentum number~\cite{deWit:1988wri, Banks:1996vh}. 

From the perspective of type IIA superstring theory, DLCQ M-theory corresponds to the so-called Matrix 0-brane Theory (M0T)~\cite{Blair:2023noj, Gomis:2023eav}, which arises from a BPS decoupling limit zooming in on a background D0-brane. In flat spacetime, such a limit is defined by sending a dimensionless parameter $\omega$ to infinity in the type IIA theory with the following reparametrizations of the target-space line element $\dd s^2$\,, the RR one-form $C^{(1)}$, and the dilaton $\Phi$~\cite{Blair:2023noj, Gomis:2023eav, Blair:2024aqz}:
\begin{align} \label{eq:dstmzt}
    \dd s^2 = - \omega \, \dd t^2 + \frac{1}{\omega} \, \dd x^i \, \dd x^i\,, 
        \qquad%
    C^{(1)} = \omega^2 \, g^{-1}_\text{s} \, \dd t\,,
        \qquad%
    e^\Phi = \omega^{-\frac{3}{2}} \, g^{}_\text{s}\,,
\end{align}
where $i = 1\,, \, \cdots, 9$ and $\omega$ is related to the Lorentz factor $\gamma$ controlling the infinite boost that leads to DLCQ M-theory, with 
$\omega = \sqrt{2} \, \gamma + O(\gamma^{-1})$ at large $\gamma$ (see Section 4.1 in~\cite{Blair:2024aqz} and also~\cite{Blair:2023noj, Gomis:2023eav}). In the $\omega \rightarrow \infty$ limit, the RR one-form charge of the background D0-brane is fine-tuned to cancel its brane tension. Moreover, the target space geometry becomes non-Lorentzian, which is now equipped with a codimension-one foliation structure and a Galilean boost symmetry relating space to time. The BFSS matrix theory captures the dynamics of the light D0-brane modes in M0T. 

Next, we consider M0T compactified on a spatial circle $x^1$\,. T-dualizing M0T in $x^1$ leads to Matrix 1-brane Theory (M1T)~\cite{Blair:2023noj, Gomis:2023eav}, whose S-dual gives rise to type IIB non-relativistic superstring theory. This chain of duality transformations acting on~\eqref{eq:dstmzt} leads to the following reparametrization of type IIB superstring theory~\cite{Bergshoeff:2019pij}:
\begin{align} \label{eq:dstmtt}
    \dd s^2 = \omega^2 \, \bigl( - \dd x_0^2 + \dd x_1^2 \bigr) + \dd x^i \, \dd x^i\,, 
        \qquad%
    B^{(2)} = \omega^2 \, \dd t \wedge \dd x^1\,,
        \qquad%
    e^\Phi = \omega \, g^{}_\text{s}\,,
\end{align}
where $i = 2\,,\, \cdots, 9$ and $B^{(2)}$ is the background Kalb-Ramond potential. The above prescription is essentially the flat case of Eq.~\eqref{eq:nrstpre}. Plugging Eq.~\eqref{eq:dstmtt} into type IIB superstring theory followed by sending $\omega$ to infinity leads to type IIB non-relativistic superstring theory, which is a perturbative string theory. This $\omega \rightarrow \infty$ limit is a BPS decoupling limit that zooms in on a background fundamental string, such that the target space geometry is equipped with a codimension-two foliation structure and enjoys a stringy generalization of the Galilean boost. Although the target space is non-Lorentzian, the string worldsheet remains Lorentzian in non-relativistic string theory, which is described by the Gomis-Ooguri sigma model that we discussed in Section~\ref{sec:rgosm}. 
The D0-brane number $N$ in M0T is now dualized to be the winding number of the IIB non-relativistic string wrapped around the compact $x^1$ circle. 

The second quantization of IIB non-relativistic string theory gives rise to matrix string theory~\cite{Motl:1997th, Dijkgraaf:1997vv}, which is described by 2D $\CN = 8$ SU($N$) SYM. We review matrix string theory and its relation to non-relativistic string theory below. According to the duality web in Figure~\ref{fig:rm1}, matrix string theory is dual to the BFSS matrix theory compactified over the spatial $x^1$ circle. Focusing on the bosonic sector, and in the unit $\alpha' \sim 1$\,, matrix string theory is described by the action ($\alpha = 0,1$)
\be \label{eq:mst}
    S = - \int \dd^2 \sigma \, \tr \Bigl( \tfrac{1}{2} \, D_{\!\alpha} X^i \, D^\alpha X^i + \tfrac{1}{4} \, g^2_\text{s} \, F_{\alpha\beta} \, F^{\alpha\beta} - \tfrac{1}{4} \, g^{-2}_\text{s} \, \bigl[ X^i\,,\,X^j \bigr]^2 \Bigr) + \text{fermions}. 
\ee
In the infrared (IR) limit where the Yang-Mills coupling $g^{}_\text{YM} \sim g_\text{s}^{-1} \rightarrow \infty$\,, the string coupling tends to zero, such that we have strongly coupled SYM described by free strings. To take the limit in the last term of Eq.~\eqref{eq:mst} we first rewrite this term by introducing an  auxiliary field $\lambda_{ij}$ as follows:
\be \label{eq:lij}
    - \tfrac{1}{4} \, g^{-2}_\text{s} \, \bigl[ X^i, X^j \bigr]^2 \rightarrow \lambda^{}_{ij} \, \bigl[ X^i, X^j \bigr] + g^2_\text{s} \, \lambda^{}_{ij} \, \lambda^{ij}\,.
\ee
In the $g^{}_\text{s} \rightarrow 0$ limit, the last term in Eq.~\eqref{eq:lij} drops out and $\lambda_{ij}$ becomes a Lagrange multiplier imposing the condition $[X^i,X^j] = 0$\,, which implies that $X^i$ must be in the Cartan subalgebra of the SU($N$) gauge group and that it can be diagonalized after a gauge transformation~\cite{Harvey:1995tg}. Traversing along the spatial circle in matrix string theory in general interchanges the eigenvalues $x^i_r$\,, $r = 1\,,\,\cdots, \, N$ of $X^i$. Hence, we are led to an $S_N$ orbifold conformal field theory in the deep IR, where $S_N$ is the Weyl group acting adjointly on $x^i$. The Hilbert space is decomposed into the so-called twisted sectors, which are labeled by the conjugacy classes of $S_N$ and each of them corresponds to a non-relativistic string with a certain length. In particular, the number of strings is conserved at $g^{}_\text{s} = 0$\,. Going away from this free-string theory implies that we introduce certain field-theoretical operators that join or split the strings. This means that matrix string theory at finite $g^{}_\text{s}$ is a second quantization of non-relativistic strings. See also further discussions in~\cite{Blair:2024aqz, Harmark:2025ikv}.  

Finally, note that dualizing the $x^1$ circle in type IIB non-relativistic string theory leads to type IIA superstring theory in the DLCQ, where the winding number of the IIB non-relativistic string is mapped to the Kaluza-Klein number in the lightlike circle~\cite{Gomis:2000bd, Danielsson:2000gi, Bergshoeff:2018yvt}. This DLCQ IIA model can be obtained alternatively from compactifying DLCQ M-theory over an extra spatial circle, which we take to be the M-theory circle. 

\subsection{Decoupling Limits of Heterotic Strings} \label{sec:dlhs}

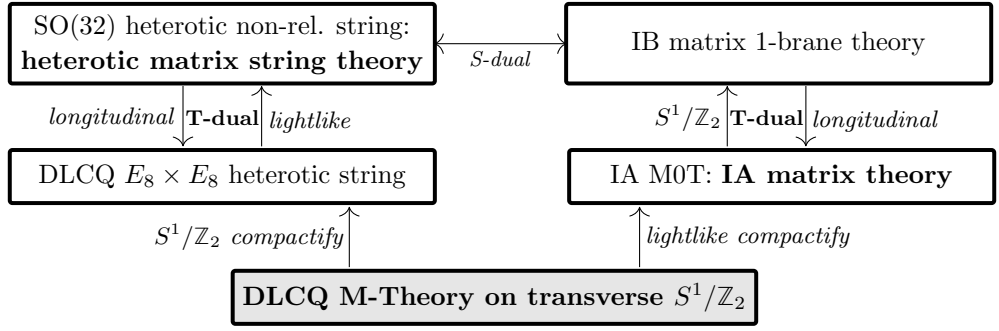
\begin{figure}[t!]
\centering
\begin{tikzpicture}
    \begin{scope}[scale=.8]

        \path[every node/.style=draw, rounded corners=1, line width=1.5pt, minimum width=160pt, minimum height=20pt, font = \small]   
                (2.5,1.5) node {DLCQ $E_8 \times E_8$ heterotic string}
                (7,-.5) node[fill=gray!20] {\textbf{DLCQ M-Theory on transverse $S^1/\mathbb{Z}_2$}}
                (11.75,1.5) node {IA M0T: \textbf{IA matrix theory}}
                (2.5,3.7) [align=center] node[minimum height = 30pt] {SO(32) heterotic non-rel. string:\\
                \textbf{heterotic matrix string theory}};

            \path[every node/.style=draw, rounded corners=1, line width=1.5pt, minimum width=160pt, minimum height=30pt, font = \small]
                (11.7,3.7) node {IB matrix 1-brane theory};

        \path   (7.1,3.25+0.2) [font=\footnotesize] node {\scalebox{0.9}{\emph{S-dual}}}
                (11+0.5,2.47) [font=\footnotesize] node {\scalebox{0.9}{\textbf{T-dual}}}
                (1.35-0.68,2.47) [font=\footnotesize] node {\emph{longitudinal}}
                (4.45-0.5,2.41) [font=\footnotesize] node {\emph{lightlike}}
                (9.35+0.85,2.47) [font=\footnotesize] node {$S^1/\mathbb{Z}_2$}
                (12.35+0.95,2.415) [font=\footnotesize] node {\emph{longitudinal}}
                (3-0.5,2.47) [font=\footnotesize] node {\scalebox{0.9}{\textbf{T-dual}}}
                (3.5-0.55,0.5) 
                [font=\footnotesize] node {\emph{$S^1/\mathbb{Z}_2$ compactify}}
                (10.3+0.9,0.5) [font=\footnotesize] node {\emph{lightlike compactify}}
               ;

         \draw[{<[length=1.3mm]}-{>[length=1.3mm]}] (6,3.5+0.2) -- (8.2,3.5+0.2);
        \draw[{[length=1.3mm]}-{>[length=1.3mm]}] (8.5+0.9,0) -- (8.5+0.9,1);
        \draw[{[length=1.3mm]}-{>[length=1.3mm]}] (5.5-0.9,0) -- (5.5-0.9,1);
       
        \draw [{<[length=1.3mm]}-{[length=1.3mm]}] (2.35-0.5,2) -- (2.35-0.5,3);
        \draw [{[length=1.3mm]}-{>[length=1.3mm]}] (3.65-0.5,2) -- (3.65-0.5,3);
        \draw [{[length=1.3mm]}-{>[length=1.3mm]}] (10.35+0.5,2) -- (10.35+0.5,3);
        \draw [{<[length=1.3mm]}-{[length=1.3mm]}] (11.65+0.5,2) -- (11.65+0.5,3);
       
\end{scope}
\end{tikzpicture}
\caption{Relations among type IA matrix quantum mechanics, heterotic matrix string theory, heterotic non-relativistic string theory, and the DLCQ of heterotic string theory.}
\label{fig:rm2}
\end{figure}

Now, we are ready to describe the heterotic analog of matrix (string) theory in type II superstring theory, following the extensive studies in the previous works~\cite{Danielsson:1996es, Kachru:1996nd, Motl:1996xx, Kim:1997uv, Lowe:1997fc, Banks:1997it, Banks:1997zs, Lowe:1997sx, Rey:1997hj, Horava:1997ns, Motl:1997tb, Krogh:1998vb, Krogh:1998rw}. Our starting point is heterotic M-theory, \emph{i.e.}~M-theory on $S^1 / \mathbb{Z}_2$ with two end-of-the-world nine-branes~\cite{Horava:1995qa, Horava:1996ma}. Each of these boundaries carries the gauge fields realizing an $E_8$ group. We furthermore consider the DLCQ of heterotic M-theory by compactifying it over a lightlike circle. In the limit where the distance between the two end-of-the-world branes in heterotic M-theory is taken to be infinite, we should recover the DLCQ M-theory in Section~\ref{eq:mtttss}. However, we have seen in Section~\ref{sec:nonhstdlcq} that the meaning of  DLCQ in the heterotic case is more subtle than in the type II case: it is the geometry associated with the modified metric $G_\text{MN}$ in~\eqref{effectivem}, instead of the actual spacetime metric $g^{}_\text{MN}$\,, that has to develop a lightlike isometry.

Analogous to the argument reviewed in Section~\ref{eq:mtttss}, we know that the dynamics of heterotic M-theory in the DLCQ is captured by the Kaluza-Klein excitations in the lightlike circle, which are the D0-branes in a corner of type IA (\emph{i.e.}~type I') superstring theory. We refer to this corner as type IA M0T, as it should arise from a BPS decoupling limit that resembles the M0T limit defined in Eq.~\eqref{eq:dstmzt}. The effective field theory of $N$ coincident D0-branes in type IA M0T is the type IA matrix theory. The conjecture analogous to the  BFSS conjecture then implies that the large $N$ limit of type IA matrix theory leads to heterotic M-theory in flat spacetime.

Next, we T-dualize type IA M0T along the interval $S^1/\mathbb{Z}_2$\,, which should lead to the type IB (\emph{i.e.}~type I) version of M1T, compactified on a dual circle (instead of an interval). Finally, we perform an S-duality transformation of type IB M1T, which supposedly gives rise to the heterotic SO(32) version of non-relativistic string theory. The formulation of the worldsheet theory of heterotic non-relativistic string theory has been the main theme throughout this paper. The second quantization of heterotic SO(32) non-relativistic string theory is heterotic matrix string theory, which is described by 2D ($0, 8$) O($N$) SYM. In the IR, the gauge theory is conjectured to flow to the orbifold CFT with $\mathbb{R}^8 / (S_N \ltimes ( \mathbb{Z}_2 )^N )$ (see \emph{e.g.}~\cite{Lowe:1997sx}).

Analogous to the T-dual relation between type IIB non-relativistic string theory and type IIA superstring theory in the DLCQ, we expect that further T-dualizing heterotic SO(32) non-relativistic string theory along the spatial longitudinal compactification defines the $E_8 \times E_8$ heterotic string in the DLCQ. This T-dual relation is what we have examined in Section~\ref{sec:hnrssm}, where the precise meaning of such a heterotic DLCQ was given. 

One subtlety to note here is that anomaly cancellation in heterotic matrix string theory requires additional chiral fermions. In the type IB language, this implies that a background of 32 D9-branes has been included.\footnote{In the conventional case before any BPS decoupling limits are taken, these spacetime-filling branes naturally occur in the  construction
of  string theories with sixteen supercharges by modding out a type II
string theory with 32 supercharges by an appropriate discrete symmetry \cite{Bergshoeff:1998re}.} Furthermore, type IB perturbation theory typically breaks down unless a specific Wilson line is introduced~\cite{Polchinski:1995df}, which corresponds in type IA to placing 8 D8-branes on each of the orientifold planes at the ends of $S^1/\mathbb{Z}_2$\,. This breaks the SO(32) gauge symmetry to SO(16)$\times$SO(16). Note that the D-branes here are the ones after performing the relevant BPS decoupling limits. In order to restore the $E_8 \times E_8$ symmetry in heterotic M-theory, \emph{i.e.}~in the limit of infinite type IA coupling, another Wilson line along $S^1$ in type IA has to be added.  

\subsection{Comments on Holographic Duality}

Finally, we comment on the holographic dual of heterotic matrix string theory. We will use the technique developed in~\cite{Blair:2024aqz}, where the bulk geometry in the holographic dual is generated by deforming the non-Lorentzian geometry at the asymptotic infinity via a D-brane generalization of the $T\bar{T}$ deformation. In the heterotic setup, it is the current-current deformation in Section~\ref{sec:ccdhst} that we apply to generate the bulk geometry. The discussions in this section follow directly the recent developments in~\cite{Blair:2024aqz}. Closely related arguments regarding the holographic setup in the type II case can be found in~\cite{Danielsson:2000mu, Avila:2023aey, Guijosa:2023qym, Harmark:2025ikv}, where the holographic dual of matrix string theory, \emph{i.e.}~the second quantization of non-relativistic string theory, is discussed.  

We return to the heterotic non-relativistic string action~\eqref{eq:nrhst123} but now in flat spacetime, with $\tau^{}_\text{M}{}^0 = \delta_\text{M}^0$\,, $\tau^{}_\text{M}{}^1 = \delta_\text{M}^1$\,, and $E^{}_\text{M}{}^i = \delta_\text{M}^i$\,, together with zero gauge potentials, \emph{i.e.}~$b^{}_\text{MN} = u^{}_\text{M} = v^{}_+ = 0$\,. We then turn on the current-current deformation~\eqref{eq:ccdef} with the background field $\mathbf{t} (X^\text{M})$\,. The deformed action is
\be \label{eq:ccdbg}
    S = - \frac{1}{4\pi\alpha'} \int \dd^2 \sigma \, \Bigl[ \p^{}_z X^i \, \p^{}_{\bar{z}} X^i + \tr \bigl( \lambda \, \p^{}_{\bar{z}} \lambda \bigr) + \chi^{}_+ \, \p^{}_{\bar{z}} X^+ + \chi^{}_- \, \p^{}_z X^- + \mathbf{t} (X) \, \chi^{}_+ \, \chi^{}_- \Bigr]\,. 
\ee
In comparison with Eq.~\eqref{eq:nrhst1def}, we have replaced the constant parameter $\omega^{-2}$ with a general background field $\mathbf{t} (X^\text{M})$\,, by choosing a different gauge of the target space dilatational symmetry. Moreover, we take the dilaton background field to be constant and denote the string coupling as $g^{}_\text{s}$\,. Also recall that $i = 2\,,\,\cdots,\,9$ and that the string wraps around the spacelike compact direction $x^1 = \frac{1}{2} (x^+ - x^-)$\,. Analogous to Eq.~\eqref{eq:nrstpre}, the deformed ten-dimensional target space geometry is then Lorentzian, with the NSNS sector being described by
\be \label{eq:tnhfs}
    \dd s^2 = \frac{1}{\mathbf{t}} \Bigl( - \dd x_0^2 + \dd x_1^2 \Bigr) + \dd r^2 + r^2 \, \dd\Omega_7^2\,,
        \qquad%
    B^{(2)} = - \frac{1}{\mathbf{t}} \, \dd x^0 \wedge \dd x^1\,,
        \qquad%
    e^\Phi = \frac{g^{}_\text{s}}{\sqrt{\mathbf{t}}}\,,
\ee
where $r^2 = x^i \, x^i$ and $x^0 = \frac{1}{2} (x^+ + x^-)$\,. 
In order to satisfy the target space supergravity equations of motion, $\mathbf{t}$ has to be the harmonic function $c + (\ell/r)^6$\,, with a constant $c$ and characteristic length scale $\ell$~\cite{Dabholkar:1995nc}. Returning to the sigma model~\eqref{eq:ccdbg}, we have learned that heterotic non-relativistic string theory, \emph{i.e.}~heterotic matrix string theory, arises in the $\mathbf{t} \rightarrow 0$ limit. In the context of holography, according to~\cite{Blair:2024aqz}, this is the BPS decoupling limit that coincides with the asymptotic limit of the bulk geometry~\eqref{eq:tnhfs} where $r / \ell \rightarrow \infty$\,. 
This requires that the constant $c$ in $\mathbf{t}$ is zero, \emph{i.e.}~\cite{Guijosa:2023qym, Harmark:2025ikv},
\be
    \mathbf{t} = \bigl( \ell / r \bigr)^6\,. 
\ee
We are thus led to the bulk geometry~\eqref{eq:tnhfs} in the holographic setup that arises from backreacting $N$ heterotic non-relativistic strings, which sets a starting point for constructing the gravity dual to heterotic matrix string theory. Analogous to the discussions in~\cite{Blair:2024aqz}, the above procedure generates the bulk Lorentzian geometry from the asymptotic non-Lorentzian geometry by turning on the current-current deformation of heterotic non-relativistic string theory. 

From the target space perspective, the geometry~\eqref{eq:tnhfs} arises from the near-horizon limit (\emph{i.e.}~the Maldacena limit) of the general string soliton background~\cite{Dabholkar:1995nc}
\be \label{eq:tnhfst}
    \dd s^2 = \frac{1}{H} \, \Bigl( - \dd \hat{t}^2 + \dd \hat{x}_1^2 \Bigr) + \dd \hat{r}^2 + \hat{r}^2 \, \dd\Omega_7^2\,,
        \qquad%
    B^{(2)} = - \frac{1}{H} \, \dd \hat{t} \wedge \dd \hat{x}^1\,,
        \qquad%
    e^\Phi = \frac{\hat{g}^{}_\text{s}}{\sqrt{H}}\,,
\ee
where the harmonic function $H$ is given by
\be
    H = 1 + \lr \frac{\hat{\ell}}{\hat{r}} \rr^{\!\!6},
        \qquad%
    \hat{l}^{\,6} = 32 \, \pi^2 \, w \, \hat{g}^2_\text{s} \, \ell^{3}_\text{s}\,,
\ee
with $\hat{r}^2 = \hat{x}^i \, \hat{x}^i$\,,  $\hat g_\text{s}$ the asymptotic string coupling at $\hat{r} \rightarrow \infty$\,, and $w$ the string winding number around the $\hat{x}^1$ circle. We have introduced the hatted notation to distinguish from the near-horizon geometry~\eqref{eq:tnhfs}. 
Following~\cite{Blair:2024aqz}, we take the BPS decoupling limit applied to the asymptotic infinity regime, 
which is prescribed by the non-relativistic string parametrization~\eqref{eq:nrstpre} (together with the dilaton reparametrization). In terms of the string soliton notation~\eqref{eq:tnhfst}, we have
\be \label{eq:nrstpret}
    \hat{t} = \omega \, t\,,
        \qquad%
    \hat{x}^1 = \omega \, x^1\,,
        \qquad%
    \hat{x}^i = x^i\,,
        \qquad%
    \hat{g}^{}_\text{s} = \omega \, g^{}_\text{s}\,.
\ee
It then follows that $\hat{\ell}^{\,6} = \omega^2 \, \ell^{\,6}$\,, with $\ell^{\,6} = 32 \, \pi^2 \, w \, g^2_\text{s} \, \ell^3_\text{s}$\,. 
Plugging Eq.~\eqref{eq:nrstpret} into the string soliton solution~\eqref{eq:tnhfst}, followed by the $\omega \rightarrow \infty$ limit, we recover the near-horizon geometry~\eqref{eq:tnhfs}~\cite{Guijosa:2023qym, Harmark:2025ikv}. This bulk semi-classical geometric description~\eqref{eq:tnhfs} is valid when the associated curvature length scale is much larger than the string length and when the string coupling is very weak, \emph{i.e.}~$1 \ll r / \ell^{}_\text{s} \ll w^{1/6}$~\cite{Harmark:2025ikv}. In this way, the BPS decoupling limit performed in the asymptotic infinity regime corresponds to the near-horizon limit in the bulk. %

The above discussions do not actually distinguish between the heterotic and type II cases. This is the universal part of the story. However, we also know that the self-consistency of heterotic matrix string theory requires introducing Wilson lines and specific brane configurations. This makes the precise holographic dual much trickier and deserves further studies. In general, we expect that the YM gauge fields do not vanish, which introduces interesting twists to the bulk geometry. For example, it is consistent to turn on constant gauge components $v_+$ and $u_\text{M}$\,, which does not introduce any non-trivial fluxes and thus the supergravity equations of motion must continue to hold. However, applying the same current-current deformation as in Eq.~\eqref{eq:ccdbg} with $\mathbf{t} = (\ell / r)^6$\,, we are now generating a different bulk geometry that is in form the same as Eq.~\eqref{eq:repa0}, but now with $\omega^{-2}$ there replaced with $\mathbf{t}$\,. It would be interesting to check whether this procedure leads to a self-consistent solution of 10D heterotic supergravity. 

\section{Conclusions} \label{sec:concl}

In this paper we studied the path integral for heterotic string theory that is anomaly free with respect to the lowest-order quantum corrections. We then used this worldsheet formalism to provide a simple derivation of the Buscher rules in heterotic string theory, in complement to the original supergravity derivation. With these techniques in hand, we studied the DLCQ of heterotic string theory and used it to motivate the worldsheet formalism for the heterotic non-relativistic string theory, whose second quantization gives rise to the heterotic matrix string theory. After establishing the new superspace  sigma models in Eq.~\eqref{eq:SS2}, we studied its YM gauge and gravitational anomalies and proposed a path integral which is anomaly free with respect to the quantum corrections at the lowest order in $\alpha'$. 

We also studied the current-current deformation under which heterotic non-relativistic string theory flows to the conventional heterotic string theory. This marginal deformation is closely related to the idea of the $T\bar{T}$-deformation, and it allowed us to systematically derive the BPS decoupling limit under which heterotic non-relativistic string theory arises. This limiting procedure is consistent with the supergravity construction in~\cite{Bergshoeff:2023fcf}.

There is still a further technical question that remains. Consider the conventional heterotic string sigma model with the modified metric~\eqref{eq:mmvvoo}, whose path integral is free of both YM gauge and gravitational anomalies with respect to the lowest-order $\alpha'$-corrections. Applying the limiting procedure that we have developed in Section~\ref{sec:ccdbps} should lead to the heterotic non-relativistic string sigma model with the modified metric~\eqref{eq:mmvvoonr}. However, under the limiting prescription, the spin connections in the modified metric~\eqref{eq:mmvvoo} contain various divergences in  $\omega$\,, which need to be taken care of. On the other hand, one should now also include $R^2$ terms in the heterotic supergravity action, which may require us to deal with (cancellations between) additional divergences. 
We leave these for future studies. 

One natural followup project is to compute the $\beta$-functions of heterotic non-relativistic string sigma models by generalizing the methods developed in~\cite{Gomis:2019zyu, Gallegos:2019icg, Bergshoeff:2019pij, Yan:2019xsf, Yan:2021lbe}, using both the CFT techniques and limiting procedure. This will reveal the target space non-Lorentzian supergravity dynamics in the presence of the YM fields. Moreover, it would be interesting to understand how the non-Lorentzian version of heterotic supergravity in~\cite{Bergshoeff:2024nin} is related to the worldsheet $\beta$-function calculation, where the latter should contain the Poisson equation.   

Our work forms an initial step towards the program of mapping out the duality web of non-perturbative corners in heterotic and type I string theory and M-theory on $S^1 / \mathbb{Z}_2$\,. It would be interesting to apply the recently developed intuitions and techniques involving non-Lorentzian geometry to understand the dynamics in this duality web, which is central to the heterotic/type I versions of matrix theory. In particular, it would be interesting to consider the BFSS conjecture in the heterotic case. At finite $N$, one expects that the type I matrix quantum mechanics computes instantaneous interactions between D0-branes in non-Lorentzian supergravity incorporating the torsional constraints $\nabla^{}_{\![\text{M}} \tau^{}_{\text{N}]}{}^+ = 0$\,. Moreover, it would be interesting to develop the heterotic version of non-Lorentzian bootstrap proposed in~\cite{Bergshoeff:2023ogz}, where non-Lorentzian bosonic symmetries are used to constrain the `parent' Lorentzian theory, by requiring the self-consistency of the BPS decoupling limit. This may help us fix higher-derivative curvature terms in heterotic supergravity without resorting to fermions. 

It would also be interesting to consider timelike T-duality of type IA M0T (see Figure~\ref{fig:rm2}). Analogous to the discussions in~\cite{Gomis:2023eav}, this procedure can be used to define a type IB version of tensionless string theory (and ambitwistor string theory in connection to the CHY formalism). Finally, it may be possible to define a second DLCQ in type IB M1T (see Figure~\ref{fig:rm2}). Following~\cite{Blair:2023noj, Gomis:2023eav}, we expect that in a T-dual frame this further DLCQ is mapped to a spatial circle in a `multicritical' type IA theory. However, novel twists are expected: as we have noted in Appendix~\ref{app:tdhnrst}, in the S-dual frame described by heterotic non-relativistic string theory, there are two different DLCQs that one could introduce. Finally, it would be interesting to consider possible holographic duals that arise from the BPS decoupling limits of heterotic and type I superstring theory that we have discussed here. 

\acknowledgments

We would like to thank Chris Blair, Troels Harmark, Johannes Lahnsteiner, and Niels Obers for useful discussions. The work of E.B. is supported by the Croatian Science Foundation project IP-2022-10-5980 “Non-relativistic supergravity and applications''. The work of L.R. has been supported in part by the Ramon y Cajal
fellowship RYC2023-042671-I , funded by
MCIU/AEI/10.13039/501100011033 and FSE+ and in part by
the MCI, AEI, FEDER (UE) grant PID2021-125700NAC22. The work of Z.Y. is
supported in part by Olle Engkvists Stiftelse Project Grant 234-0342, VR Project Grant 2021-04013, and the European Union’s Horizon 2020 research and innovation programme
under the Marie Sklodowska-Curie Grant Agreement No.~31003710.

\appendix

\section{Stueckelberg Symmetry} \label{app:ss}

It is curious that the sigma model~\eqref{eq:dgeim} describing the heterotic non-relativistic string acquires an extra component $v^{}_+$ in addition to the gauge potential $u^{}_\text{M}$\,. This is because one component of $u^{}_\text{M}$ acts as a Stueckelberg field and can be eliminated by redefining $\chi^{}_+$ at the classical level. In fact, the action~\eqref{eq:nrhst123} is invariant under the Stueckelberg transformation,
\be
    \chi^{}_+ \rightarrow \chi^{}_+ - i \, \tr \Bigl( \lambda \, \CC^{}_+ \, \lambda \Bigr)\,,
        \qquad%
    u^{}_\text{M} \rightarrow u^{}_\text{M} - \CC^{}_+ \, \tau^{}_\text{M}{}^+\,,
\ee
for an arbitrary function $\CC_+$\,.
In order to fix this Stueckelberg symmetry, it is useful to define the inverse vielbein fields $\tau^\text{M}{}^{}_I$ and $E^\text{M}{}^{}_i$ via the orthogonality conditions,
\begin{subequations}
\begin{align}
    \tau^\text{M}{}^{}_I \, \tau^{}_\text{M}{}^J & = \delta^J_I\,, 
        &%
    E^\text{M}{}^{}_{i} \, \tau^{}_\text{M}{}^I = \tau^\text{M}{}^{}_{I} \, E^{}_\text{M}{}^{i} & = 0\,, \\[4pt]
    E^\text{M}{}^{}_{i} \, E^{}_\text{M}{}^{j} & = \delta^{j}_{i}\,,
        &%
    \tau^{}_\text{M}{}^I \, \tau^\text{N}{}_{I} + E^{}_\text{M}{}^{i} \, E^\text{N}{}_{i} & = \delta_{\text{M}}^{\text{N}}\,. 
\end{align}
\end{subequations}
Decompose $u^{}_\text{M}$ as
$u^{}_\text{M} = \tau^{}_\text{M}{}^+ \, u_+ + \tau^{}_\text{M}{}^- \, u_- + E^{}_\text{M}{}^i \, u_i$\,,
with $u^{}_\pm \equiv u^{}_\text{M} \, \tau^\text{M}{}^{}_\pm$ and $u^{}_i \equiv u^{}_\text{M} \, E^\text{M}{}^{}_i$\,. 
Setting $\CC^{}_+ = u^{}_+$\,, 
the heterotic non-relativistic string action~\eqref{eq:nrhst123} becomes
\begin{align} \label{eq:nrhst1}
\begin{split}
	S = \frac{1}{4\pi} \int \dd^2 z \, \biggl\{ &\p^{}_z X^\text{M} \, \p^{}_{\bar{z}} X^\text{N} \, \Bigl( E^{}_{\text{M}}{}^i \, E^{}_{\text{N}}{}^i + b^{}_\text{MN} \Bigr) + \tr \Bigl( \lambda \, \p^{}_{\bar{z}} \lambda - i \, \lambda \, U^{}_{\text{M}} \, \p^{}_{\bar{z}} X^\text{M} \, \lambda \Bigr) \\[4pt]
    &\qquad\qquad\,\,\,\, + \chi^{}_+ \, \p^{}_{\bar{z}} X^{\text{M}} \, \tau^{}_{\text{M}}{}^+ + {\chi}^{}_- \Bigl[ \p^{}_z X^\text{M} \, {\tau}^{}_{\text{M}}{}^- + i \, \tr \bigl( \lambda \, v^{}_+ \, \lambda \bigr) \Bigr]  \biggr\}\,, 
\end{split}
\end{align}
where
$U^{}_\text{M} = \tau^{}_\text{M}{}^- \, u^{}_- + E^{}_\text{M}{}^i \, u_i$\,. We continue to use the notation $\chi^{}_+$ even though it has been redefined. 
Note that the $u^{}_+$ component is now removed, such that we have the usual number of components in the gauge multiplet $(v^{}_+\,,\,u^{}_-\,,\,u^{}_i)$\,. 
In this different representation, the gauge transformation~\eqref{eq:gthsnr} becomes
\begin{subequations} \label{eq:gtluvpcp}
\begin{align}
    \delta_\xi \lambda &= i \, \xi \, \lambda\,, 
        &%
    \delta_\xi U^{}_\text{M} &= \tau^{}_\text{M}{}^- \, \nabla_{\!-} \xi + E^{}_\text{M}{}^i \, \nabla_{\!i} \xi\,, \\[4pt]
    \delta_\xi v^{}_+ &= i \bigl[\xi\,,\,v^{}_+\bigr]\,,
        &%
    \delta_\xi \chi^{}_+ &= - i \, \tr \bigl( \lambda \, \nabla^{}_{\!+} \xi \, \lambda \bigr)\,.
\end{align}
\end{subequations}
However, this formulation is not convenient for the anomaly analysis. We therefore focused on the original formulation~\eqref{eq:nrhst123} in the bulk of the paper. 

\section{T-Duality in Heterotic Non-Relativistic String Theory} \label{app:tdhnrst}

At the end of Section~\ref{sec:hnrssm}, we considered the T-duality transformation along a longitudinal spatial isometry in heterotic non-relativistic string sigma models, where the T-dual frame is described by the DLCQ of the conventional heterotic string. In this appendix, we generalize~\cite{Bergshoeff:2018yvt} to study other types of T-duality transformations in heterotic non-relativistic string sigma models, including ones along a transverse spatial isometry and along longitudinal lightlike isometries. 

We start with a transverse T-duality transformation in an isometry $x$ satisfying
$\tau^{}_x{}^A = 0$ and $E^{}_{xx} \neq 0$\,.
The T-dual action is
\begin{align} \label{eq:nrhst12}
\begin{split}
	\tilde{S} = - \frac{1}{4\pi\alpha'} \int \dd^2 \sigma \, \biggl\{ \p^{}_z X^\text{M} \, \p^{}_{\bar{z}} X^\text{N} \, \tilde{\CE}^{}_{\text{M}\text{N}} &+ \tr \Bigl( \lambda \, \p^{}_{\bar{z}} \lambda - i \, \lambda \, \tilde{U}^{}_{\text{M}} \, \p^{}_{\bar{z}} X^\text{M} \, \lambda \Bigr) \\[4pt]
    + \chi^{}_+ \, \p^{}_{\bar{z}} X^{\text{M}} \, \tau^{}_{\text{M}}{}^+ &+ {\chi}^{}_- \, \Bigl[ \p^{}_z X^\text{M} \, {\tau}^{}_{\text{M}}{}^- + i \, \tr \bigl( \lambda \, v^{}_+ \, \lambda \bigr) \Bigr]  \biggr\}\,, 
\end{split}
\end{align}
where the Buscher rules are
\begin{subequations} \label{eq:hetbuscher0}
\begin{align}
    \tilde{\CE}_{xx} &= \frac{1}{\CE_{xx}}\,,
        &
    \tilde{\CE}_{x \mu} &= \frac{\CE_{x \mu}}{\CE_{xx}}\,,
        &
    \tilde{U}_{x} &= - \frac{U_{x}}{\CE_{xx}}\,, \\[4pt]
    \tilde{\CE}_{\mu\nu} &= \CE_{\mu\nu} - \frac{\CE_{\mu x} \, \CE_{x \nu}}{\CE_{xx}}\,,
        &
    \tilde{\CE}_{\mu x} &= - \frac{\CE_{\mu x}}{\CE_{xx}}\,,
        &
    \tilde{U}_{\mu} &= U_{\mu} - \frac{U_x \, \CE_{x \mu}}{\CE_{xx}}\,.
\end{align}
\end{subequations}
Moreover, the dilaton $\phi$ transforms into
$\tilde{\phi} = \phi - \frac{1}{2} \ln \CE_{xx}$\,. 
This dual action still describes the heterotic non-relativistic string.

It is also possible to consider the DLCQ of heterotic non-relativistic string theory and perform a T-duality transformation along the associated lightlike direction. However, there are two different possibilities in this heterotic case. First, we consider an isometry $x$ satisfying
$\tau^{}_x{}^+ \neq 0$\,, $\tau^{}_x{}^- = 0$\,, and $E^{}_x{}^i = 0$\,.
In order to obtain the T-dual action in the desired form, we first re-express the term
\be \label{eq:cpt}
    \chi^{}_+ \, \p^{}_{\bar{z}} X^\text{M} \, \tau^{}_\text{M}{}^+
\ee
in the original action~\eqref{eq:nrhst1} equivalently as (see a similar trick in~\cite{Bergshoeff:2018yvt})
\be \label{eq:rew}
    \Bigl( f \, \p_{\bar{z}} X^\text{M} - \bar{f} \, \p_z X^\text{M} \Bigr) \, \tau^{}_\text{M}{}^+ + \tilde{\chi}^{}_+ \, \Bigl( e \, \bar{f} - \bar{e} \, f \Bigr)\,,
\ee
where we have introduced a series of auxiliary fields. Note that integrating out the Lagrange multiplier $\tilde{\chi}_+$ imposes the constraint $e \, \bar{f} = \bar{e} \, f$\,, which is solved by $f = \chi_+ \, e$ and $\bar{f} = \chi_+ \, \bar{e}$\,. Plugging these solutions back into Eq.~\eqref{eq:rew} followed by fixing $e = 1$ and $\bar{e} = 0$ gives back the original expression~\eqref{eq:cpt}. Instead, we now perform the T-duality transformation of the re-expressed action, followed by integrating out $f$ and $\bar{f}$ and then setting $e = 1$ and $\bar{e} = 0$\,. The above procedures finally give rise to the T-dual action below: 
\begin{align} \label{eq:nrhst13}
\begin{split}
	\tilde{S} = - \frac{1}{4\pi\alpha'} \int \dd^2 \sigma \, \biggl\{ \p^{}_z X^\text{M} \, \p^{}_{\bar{z}} X^\text{N} \, \tilde{\CE}^{}_{\text{M}\text{N}} &+ \tr \Bigl( \lambda \, \p^{}_{\bar{z}} \lambda - i \, \lambda \, U^{}_{\text{M}} \, \p^{}_{\bar{z}} X^\text{M} \, \lambda \Bigr) \\[4pt]
    + \tilde{\chi}^{}_+ \, \p^{}_{\bar{z}} X^{\text{M}} \, \tilde{\tau}^{}_{\text{M}}{}^+ &+ \chi^{}_- \, \Bigl[ \p^{}_z X^\text{M} \, {\tau}^{}_{\text{M}}{}^- + i \, \tr \bigl( \lambda \, v^{}_+ \, \lambda \bigr) \Bigr]  \biggr\}\,, 
\end{split}
\end{align}
where the associated Buscher rules are
\begin{subequations}
\begin{align}
    \tilde{\tau}_x{}^+ &= \frac{1}{\tau_x{}^+}\,,
        &%
    \tilde{\CE}_{xx} &= 0\,,
        \qquad%
    \tilde{\CE}_{x\mu} = - \tilde{\CE}_{\mu x} = \frac{\tau_\mu{}^+}{\tau_x{}^+}\,, \\[4pt]
    \tilde{\tau}_\mu{}^+ &= \frac{\CE_{x\mu} \, \tau_x{}^+ - \CE_{xx} \, \tau_\mu{}^+}{(\tau_x{}^+)^2}\,,
        &%
    \tilde{\CE}_{\mu\nu} &= \CE_{\mu\nu} + \frac{\tau_\mu{}^+ \, \tau_\nu{}^+}{(\tau_x{}^+)^2} \, \CE_{xx} - \frac{\tau_\mu{}^+ \, \CE_{x\nu} + \tau_\nu{}^+ \, \CE_{\mu x}}{\tau_x{}^+}\,.
\end{align}
\end{subequations}
The dilaton $\phi$ transforms into
$\tilde{\phi} = \phi - \ln \bigl| \tau_x{}^+ \bigr|$\,.
The dual theory still describes the DLCQ of the heterotic non-relativistic string. 

Alternatively, it is also possible to instead consider the isometry satisfying
$\tau^{}_x{}^+ = 0$\,,
$\tau^{}_x{}^- \neq 0$\,,
and
$E^{}_x{}^i = 0$\,.
In order to perform the T-dual transformation, we start with a similar rewriting as we did in Eq.~\eqref{eq:rew}, but now for the term containing $\chi_-$ in the heterotic non-relativistic string action~\eqref{eq:nrhst1}. Namely, we write 
\be \label{eq:cmtw}
    {\chi}^{}_- \, \Bigl[ \p^{}_z X^\text{M} \, {\tau}^{}_{\text{M}}{}^- + i \, \tr \bigl( \lambda \, v^{}_+ \, \lambda \bigr) \Bigr]
\ee
equivalently as
\be \label{eq:rewt}
    f \,\p^{}_{\bar{z}} X^\text{M} \, \tau^{}_\text{M}{}^-  - \bar{f} \, \Bigl[ \p^{}_z X^\text{M} \, {\tau}^{}_{\text{M}}{}^- + i \, \tr \bigl( \lambda \, v^{}_+ \, \lambda \bigr) \Bigr] + \tilde{\chi}^{}_- \, \Bigl( e \, \bar{f} - \bar{e} \, f \Bigr)\,.
\ee
We now solve the constraint imposed by the Lagrange multiplier $\tilde{\chi}_-$ by $f = \chi_- \, e$ and $\bar{f} = \chi_- \, \bar{e}$\,. Plugging these solutions back into Eq.~\eqref{eq:rewt} followed by setting $e = 0$ and $\bar{e} = -1$ gives back the original expression~\eqref{eq:cmtw}. Instead, performing a T-duality transformation of the re-expressed action along the isometry $x$ gives 
\begin{align} \label{eq:nrhst14}
\begin{split}
	\tilde{S} = - \frac{1}{4\pi\alpha'} \int \dd^2 \sigma \, \biggl\{ \p^{}_z X^\text{M} \, \p^{}_{\bar{z}} X^\text{N} \, \tilde{\CE}^{}_{\text{M}\text{N}} &+ \tr \Bigl( \lambda \, \p^{}_{\bar{z}} \lambda - i \, \lambda \, U^{}_{\text{M}} \, \p^{}_{\bar{z}} X^\text{M} \, \lambda \Bigr) \\[4pt]
    + \chi^{}_+ \, \p^{}_{\bar{z}} X^{\text{M}} \, \tau^{}_{\text{M}}{}^+ &+ \tilde{\chi}^{}_- \, \Bigl[ \p^{}_z X^\text{M} \, \tilde{\tau}^{}_{\text{M}}{}^- + i \, \tr \bigl( \lambda \, \tilde{v}^{}_+ \, \lambda \bigr) \Bigr]  \biggr\}\,, 
\end{split}
\end{align}
with the Buscher rules
\begin{subequations}
\begin{align}
    \tilde{\tau}_{x}{}^{-} &= \frac{1}{\tau_{x}{}^{-}}\,,
        &%
    \tilde{\CE}_{xx} &= 0\,,
        \quad%
    \tilde{\CE}_{\mu x} = \frac{\tau_{\mu}{}^{-}}{\tau_{x}{}^{-}}\,,
        \quad%
    \tilde{\CE}_{x \mu} = -\frac{\tau_{\mu}{}^{-}}{\tau_{x}{}^{-}}\,, \\[4pt]
    \tilde{\tau}_{\mu}{}^{-} &= \frac{\tau_{x}{}^{-} \, \CE_{z\mu } - \tau_{\mu}{}^{-} \, \CE_{xx}}{(\tau_{x}{}^{-})^{2}}\,,
        &%
    \tilde{\CE}_{\mu\nu} &= \CE_{\mu\nu}
+\frac{\tau_{\mu}{}^{-} \, \tau_{\nu}{}^{-}}{(\tau_{x}{}^{-})^{2}} \, \CE_{xx}-\frac{\tau_{\mu}{}^{-} \, \CE_{x\nu} + \tau_{\nu}{}^{-} \, \CE_{\mu x}}{\tau_{x}{}^{-}}
\,,\\[4pt]
    \tilde{U}_{x} &= -\frac{v_{+}}{2 \, \tau_{x}{}^{-}}\,,
        &%
    \tilde{U}_{\mu} &= U_{\mu}+\frac{\tau_{x}{}^{-} \, \CE_{x\mu} - \tau_{\mu}{}^{-} \, \CE_{xx}}{2 \, (\tau_{x}{}^{-})^{2}} \, v_{+}-\frac{\tau_{\mu}{}^{-}}{\tau_{x}{}^{-}} \, U_{x}\,, \\[4pt]
\tilde{v}_{+}&=-\frac{\CE_{xx}}{(\tau_{x}{}^{-})^{2}} \, v_{+}-\frac{2 \, U_{z}}{\tau_{z}{}^{-}}\,.
\end{align}
\end{subequations}
The dilaton $\phi$ is T-dualized into
$\tilde{\phi} = \phi - \ln |\tau_x{}^-|$\,.

When the Yang-Mills field is set to zero, the Buscher rules found here reduce to the ones derived in~\cite{Bergshoeff:2018yvt}. In the type II case, the lightlike circle in the DLCQ of non-relativistic string theory is also T-dualized to be a lightlike circle. However, after an S-duality is performed, the lightlike circle T-dualizes to a spacelike circle~\cite{Blair:2023noj, Gomis:2023eav}. It would be interesting to explore the analogous story in the heterotic case, where S-duality is more complex. 

\bibliographystyle{JHEP}
\bibliography{hssm}

\end{document}